\documentclass[reprint,superscriptaddress,showpacs,preprintnumbers,amsmath,amssymb,pra]{revtex4-1}

\usepackage{graphicx}
\usepackage{dcolumn}
\usepackage{bm}
\usepackage{graphicx}
\usepackage{enumerate}
\usepackage{subfigure}
\usepackage[colorlinks,
linkcolor=red,
anchorcolor=black,
citecolor=blue]{hyperref}

\begin{document}

\title{Phenomenological Single-Particle Green's Function for the Pseudogap and Superconducting Phases of High-$T_c$ Cuprates}

\author{Jian-Hao Zhang}
\affiliation{Institute for Advanced Study, Tsinghua University, Beijing 100084, China}
\author{Sen Li}
\affiliation{Institute for Advanced Study, Tsinghua University, Beijing 100084, China}
\author{Yao Ma}
\affiliation{Department of Applied Physics, Xi’an University of Technology, Xi’an, Shannxi 710048, China}
\author{Yigui Zhong}
\affiliation{Beijing National Laboratory for Condensed Matter Physics and Institute of Physics, Chinese Academy of Sciences, Beijing 100190, China}
\affiliation{School of Physics, University of Chinese Academy of Sciences, Beijing 100190, China}
\author{Hong Ding}
\affiliation{Beijing National Laboratory for Condensed Matter Physics and Institute of Physics, Chinese Academy of Sciences, Beijing 100190, China}
\affiliation{School of Physics, University of Chinese Academy of Sciences, Beijing 100190, China}
\affiliation{CAS Center for Excellence in Topological Quantum Computation, University of Chinese Academy of Sciences, Beijing, 100190, China}
\author{Zheng-Yu Weng}
\affiliation{Institute for Advanced Study, Tsinghua University, Beijing 100084, China}

\date{Oct. 27, 2019}
\revised{\today}

\begin{abstract}
We present a phenomenological Green's function to characterize the superconducting and pseudogap phases of the cuprates based on a microscopic theory of doped Mott insulators. In this framework, the ``Fermi arc'' and ``kink'' phenomena observed by angle-resolved photoemission spectroscopy (ARPES) experiments in the pseudogap phase can be systematically explained as a function of doping, which are further connected to the two-gap feature in the superconducting phase with dichotomy between the nodal and antinodal physics. We demonstrate that a phase-string-induced fractionalization plays the key role in giving rise to such a peculiar Green's function with a unique two-component structure.
\end{abstract}

\pacs{74.20.-z, 74.20.Mn, 74.72.-h, 74.25.Jb}

\maketitle
\newcommand{\lra}{\longrightarrow}
\newcommand{\xra}{\xrightarrow}
\newcommand{\ra}{\rightarrow}
\newcommand{\bs}{\boldsymbol}
\newcommand{\ul}{\underline}
\newcommand{\1}{\text{\uppercase\expandafter{\romannumeral1}}}
\newcommand{\2}{\text{\uppercase\expandafter{\romannumeral2}}}
\newcommand{\3}{\text{\uppercase\expandafter{\romannumeral3}}}
\newcommand{\4}{\text{\uppercase\expandafter{\romannumeral4}}}
\newcommand{\5}{\text{\uppercase\expandafter{\romannumeral5}}}
\newcommand{\6}{\text{\uppercase\expandafter{\romannumeral6}}}

\tableofcontents

\section{\label{section1}Introduction}
The discovery of high-$T_c$ cuprate superconductors \cite{Bednorz1986} has generated tremendous interests in the past three decades not only because of the high transition temperature of unconventional superconductivity, but also on account of numerous fascinating phenomena and intertwined orders \cite{AndersonBook,ShenZX2003,Patrick2006,intertwined,Keimer2015}.
The rich experimental observations may have generically indicated the failure of the standard Landau Fermi liquid theory \cite{Abrikosov,Baym}, as one of the greatest triumphs of the condensed matter theory in the 20th century. 

The angle-resolved photo-emission spectroscopy (ARPES)\cite{Timusk1999,ZXShen2001,ShenZX2003,ZXShen2010} has been one of the most powerful tools in the experimental investigation of the high-$T_c$ cuprate materials with essential two-dimensionality (2D). 
A striking distinction of the high-$T_c$ cuprate from a Landau Fermi liquid theory description is the ARPES observation of a ``Fermi surface" consisting of four disconnected portions known as the Fermi arcs \cite{ShenZX1996,Ding1998,ShenZX2005} in the pseudogap phase, in contrast to a conventional full Fermi surface contour which can only terminate at the Brillouin zone boundary. Upon entering the superconducting phase, the Fermi arc is further gapped by a $d$-wave symmetry gap, but outside the Fermi arc, a new quasiparticle-like peak emerges, which is correlated with a ``pseudogap'' in the antinodal regime exhibiting the ``peak-dip-hump'' structure in energy distribution curves (EDCs) \cite{Shen1991,Shen1993,Loeser1997,Norman1997,Fedolov1999,DLFeng2000,Sato2002}.  Such a two-gap structure \cite{dichotomy,Sawatzky2008,Chen2009,ZXShen2014} in the superconducting phase, with the dichotomy between the nodal and antinodal quasiparticle excitations,
is further supplemented by another ``kink'' dispersion along the nodal direction which persists up to the pseudogap phase \cite{ZXShen2000,Lanzara2001,Johnson2001,Gromko2002,XJZhou2003,Keimer2006,ZhongYG2018,ZhongYG2019}. A more detailed account of the APRES data used in the present work will be further elaborated later in the Introduction. 

The significance of the ARPES measurement is that it directly probes into the behavior of the quasiparticle-like excitation that is the basic building block of a Fermi liquid state or BCS superconducting state. Thus, any anomalies shown in the APRES may indicate whether or how a conventional Fermi liquid or BCS state breaks down \cite{ding1996spectroscopic,Ding1998}. In the case of the cuprate, as mentioned above, the ARPES data have clearly demonstrated a very rich phenomenon with an incomplete Landau/Bogoliubov-type quasiparticle at least in the underdoped and optimal doping regimes. The essential question is whether such a complex phenomenon can be still understood within a modified Fermi liquid framework or a completely new phenomenology will be needed to sensibly provide a consistent picture, which must be also in accordance with a huge number of \emph{other} constraints imposed by the experiments as well as theoretical studies. 

For instance, with reducing doping, the cuprate materials will eventually recover an insulating phase in which spins become antiferromagnetically long-range ordered (AFLRO). In particular, at half-filling, the system is a Mott insulator \cite{AndersonRVB, AndersonBook}, based on which a single hole may be created by photo-emission, as has been studied by the APRES in Sr$_2$CuO$_2$Cl$_2$ and Ca$_2$CuO$_2$Cl$_2$\cite{Wells1995, LaRosa1997, Kim1996,Kim1997,Kim1998,Ronning1998,Zhou2018}. These experiments have shown that the four Fermi arcs in the pseudogap phase of finite doping have shrunk into four Fermi points at  $(\pm \pi/2, \pm\pi/2)$ in the Brillouin zone. The Fermi point positions are in agreement with the exact diagnalization (ED) calculations based on the simplified models for doped Mott insulators like the $t$-$J$ model \cite{Dagotto,Leung1995,Leung1997}, which has been also studied by analytic methods \cite{Laughlin1997,Tohyama2000,Weng2001} like the self-consistent Born approximation \cite{TMRice1970,Varma1988,Patrick1989,Horsch1991,Manousakis1991}. Recently a combined ED and density matrix renormalization group (DMRG) study has further shown that besides the quasiparticle spectral weight peaked at $(\pm \pi/2, \pm\pi/2)$ in a finite-size system \cite{ZhengWei2018}, the single hole is accompanied by a persistent spin current hidden in the spin antiferromagnetic background, which can be characterized by a novel angular momentum for the $t$-$J$ system under an open boundary with the discrete C$_4$ rotational symmetry. In other words, in the dilute hole limit, the charge carrier in the doped Mott insulator is not a Landau-like quasiparticle and the substantial broadening of the spectral function around $(\pm \pi/2, \pm\pi/2)$ observed  \cite{Greven1994, Wells1995, LaRosa1997, Kim1996,Kim1997,Kim1998,Kim2002} in the cuprate may be simply due to such momentum-carrying spin currents generated by the motion of the hole in the background.  As a matter of fact, a one-hole ground state wavefunction with incorporating such a spin current pattern has been recently calculated by variational Monte Carlo (VMC) method \cite{SingleHole}, which reproduces the quasiparticle spectral weight, momentum distribution as well as the novel angular momenta and corresponding ground state degeneracy, in excellent agreement with the ED and DMRG results of finite sizes up to $8\times 8$ \cite{ZhengWei2018}, while further shows the vanishing quasiparticle spectral weight in the thermodynamic limit. 

At a finite doping, along the line of thinking based on the doped Mott insulator or doped antiferromagnet \cite{AndersonRVB, AndersonBook, Patrick2006,Weng2011}, a superconducting (SC) ground state with a pseudogap ``normal state'' is expected to emerge after the AFLRO is doped away. Various possible SC ground states have been proposed within the same framework described by the $t$-$J$ and Hubbard models. Among them, the most influential one is the original proposal of the Gutzwiller-projected BCS ground state, i.e., the so-called ``plain vanilla'' resonating-valence-bond (RVB) state \cite{AndersonRVB,plainvanilla} and their mean-field description \cite{Baskaran1987,Baskaran1988,ZhangFC1988}. The quasiparticle excitation in such an RVB state can be generally described by a spin-charge separation with emergent gauge field in terms of the slave-boson scheme and a tremendous investigation has been conducted \cite{Patrick2006}, in which the corresponding single-particle Green's function has been studied \cite{Patrick1998,Ng2005}. A phenomenological one has been also constructed \cite{YRZ} in comparison with the ARPES experiments, which may have some close connection with the slave-boson scheme as pointed out in Ref.  \onlinecite{Ng2005}.

Alternatively, a new superconducting/pseudogap ground state of the $t$-$J$ model has been recently proposed \cite {Weng2011,MaYao2014}, which is distinct from the ``plain vanilla'' RVB state \cite{AndersonRVB} by having a two-component structure that can be continuously connected to the AFLRO state at half-filling. In the zero doping limit, such a ground state can be naturally reduced to that of the Heisenberg Hamiltonian, $|\mathrm{RVB}\rangle$, which well describes \cite{LDA} the background AFLRO. The one-hole ground state \cite{SingleHole} is created by a \emph{twisted} hole creation operator $\tilde{c}$ by $\tilde{c}|\mathrm{RVB}\rangle$, which reproduces the spectral weight of the doped hole, spin currents, and novel quantum numbers, in excellent comparison with the above-mentioned ED and DMRG results \cite{ZhengWei2018}. Two of such holes can further form a strong pairing state $\sum_{i,j}g_{ij}\tilde{c}_{i\uparrow}\tilde{c}_{i\downarrow}|\mathrm{RVB}\rangle$ as shown \cite{TwoHole} in a two-leg ladder with a spin gap, which also well reproduces the DMRG result \cite{ZhengZhu2016}. At finite doping, these doped holes finally form a BCS-like pairing state \cite {Weng2011,MaYao2014}
\begin{align}
|\Psi_{\mathrm{G}}\rangle\propto\exp\left(\sum\limits_{i,j}g_{ij}\tilde{c}_{i\uparrow}\tilde{c}_{j\downarrow}\right)|\mathrm{RVB}\rangle
\label{ground state wavefunction1}
\end{align}
in the pseudogap phase, which becomes the true Cooper pairing or SC state, with $$\sum_{i,j}g_{ij}\tilde{c}_{i\uparrow}\tilde{c}_{i\downarrow}\rightarrow \sum_{i,j}g_{ij}'\hat{c}_{i\uparrow}\hat{c}_{i\downarrow}$$ once the phase coherence is realized in $g_{ij}'$ (see below), where the background AFLRO in $|\mathrm{RVB}\rangle$ is also self-consistently reduced to a short-range AF ordered state \cite {Weng2011,MaYao2014}.  

The key to such a two-component RVB state is a peculiar fractionalization of the bare hole creation operator given by
\begin{equation}\label{tildec}
\hat{c}_{i\sigma}=\tilde{c}_{i\sigma}e^{i\hat{\Omega}_i} ~,
\end{equation}
where the phase factor $e^{i\hat{\Omega}_i}$ is a many-body operator acting on the neutral spin background $|\mathrm{RVB}\rangle$, which determines the spin current pattern induced by the hole's motion. Here $\tilde{c}_{i\sigma}\equiv \hat{c}_{i\sigma}e^{-i\hat{\Omega}_i}$ depicts a new \emph{composite} entity with a bare hole bound to a vortex of neutral spin currents as have been carefully analyzed in the one-hole case \cite{SingleHole}. The twisted hole created by $\tilde{c}_{i\sigma}$ can generally propagate coherently on $|\mathrm{RVB}\rangle$ but $\hat{c}_{i\sigma}$ will get strongly frustrated by $e^{i\hat{\Omega}_i}$ \cite{SingleHole}. Correspondingly, the ground state wavefunction at finite doping has essentially the same form in the SC and pseudogap phases in terms of $\tilde{c}$, but the two phases are distinguished as distinct ones according to the phase (in)coherence of the phase factor $\langle e^{i\hat{\Omega}_i}\rangle \neq 0$ (or $=0$) \cite {Weng2011,MaYao2014}. 

Therefore, the ``phase fractionalization'' in Eqs. (\ref{ground state wavefunction1}) and (\ref{tildec}) is in sharp contrast to the usual spin-charge fractionalization in the slave-boson scheme for the ``plain vanilla'' RVB state \cite{Patrick2006}. One expects a drastic distinction in the predictions of the single-particle Green's function by these two different ground states. So the ARPES experiment can provide a direct probe into the nature of the ground state via the particular fractionalization effect of the quasiparticle excitation. 

In this paper, we shall explore the quasiparticle excitation in the new SC/pseudogap ground state outlined above under the fractionalization of Eq. (\ref{tildec}). Here, at the mean-field level, the composite holes described by $\tilde{c}_{i\sigma}$ will propagate coherently and occupy the \emph{Fermi pockets} commensurate with doping, which further experience a BCS-like pairing instability as shown in Eq. (\ref{ground state wavefunction1}) \cite{MaYao2014}. Since the RVB background characterized by the short-ranged AF state  $|\mathrm{RVB}\rangle$ at finite doping is also gapped, the only possible gapless excitation in the present SC/pseudogap states will be the quasiparticle excitation emerging within the gap as a \emph{collective} mode based on the fractionalization expression in Eq. (\ref{tildec}). Based on an RPA-like scheme in the fractionalized mean-field state, we construct a single-hole propagator phenomenologically. The central characteristic of this Green's function is a minimal two-component structure composed of the fractionalization component determined by Eq. (\ref{tildec}) and a conventional quasiparticle propagator as a bound state of Eq. (\ref{tildec}). 

In the pseudogap phase without the phase coherence in $e^{i\hat{\Omega}_i}$, a single hole would generally behave incoherently according to Eq. (\ref{tildec}). However, we find an emergence of partial large Fermi surface pieces (i.e., Fermi arcs) in the spectral function at frequency $\omega=0$, along which a ``sharp'' quasiparticle peak is still present. Each Fermi arc coincides roughly with the inner portion of the large bare Fermi surface that is intercepted by the Fermi pockets of the fractional fermions of $\tilde{c}$ at low doping. Physically, it means that the bare hole is forbidden to decay into the more coherent fermion $\tilde{c}$ because of the Pauli exclusion principle inside the Fermi pocket of the latter that are centered at the momenta $(\pm \pi/2, \pm\pi/2)$. In other words, the ending points of the Fermi arcs correspond to the starting points of the electron fractionalization in terms of Eq. (\ref{tildec}). We also find that with the increase of doping, the pairing gap of the $\tilde{c}$ fermions is also enhanced such that the large bare-band Fermi surface near $\omega\simeq 0$ gets less truncated by the high-energy hole pockets in the overdoping.

As a unique prediction, along the diagonal direction between $(0,0)$ and $(\pm \pi/2, \pm\pi/2)$, one finds a ``kink'' feature in the quasiparticle dispersion moving away from the Fermi arcs at the termination point of the quasiparticle beyond the circles of the Fermi pockets. In particular, the theory predicts the momentum and energy scales of the kink position as a function of doping dependence, together with the ``Fermi'' velocities on the two sides of the dispersion, which are in remarkable agreement with the experiment measurement (see below). Such a feature remains unchanged in the SC phase as the $d$-wave SC gap vanishes along the diagonal direction. However, the Fermi arc will get gapped by the $d$-wave gap as the SC phase coherence is realized by $\langle e^{i\hat{\Omega}_i}\rangle\neq 0$ \cite{Weng2011,MaYao2014}. Outside the Fermi arc, a new Bogoliubov quasiparticle mode will also emerge along the large Fermi surface in the antinodal regime, which is correlated with the energy of the fractionalized fermion $\tilde{c}$ to result in a peak-dip-hump EDCs \cite{Shen1991,Shen1993,Loeser1997,Norman1997,Fedolov1999,DLFeng2000,Sato2002}.  Thus, a new ``kink'' or two-gap structure is exhibited in the SC phase along the large Fermi surfacer with a dichotomy between the nodal and antinodal regimes, where their doping dependent two-gap scales and the quasiparticle spectral weights show distinct behaviors, which are also in excellent agreement with the ARPES experiment.

Here, to systematically compare the theoretical spectral function with the experiment, we will use the APRES data taken from a prototypical high-temperature superconducting cuprate Bi$_2$Sr$_2$CaCu$_2$O$_{8+x}$ (Bi-2212) at different doping levels continuously acquired by in-situ ozone/vacuum annealing on the same sample \cite{ZhongYG2018,YGZhong2018}. After removing the uncontrollable influence of cleaving surface, the results are highly comparable and precise enough to do the quantitative analysis. The systematic in-situ APRES measurements show a \emph{strong} dichotomy between nodal and antinodal region no matter in quasiparticle excitation energy gap or quasiparticle spectral weight, and their doping dependence, which are self-consistently explained without fitting parameters by the present theory. For superconducting energy gap, the gap around node (within Fermi arc) follows with $d$-wave gap symmetry and its slope matches well with the forming energy scale of Cooper pairs determined by the Nernst effect or diamagnetism measurements \cite{YayuWang2000,YayuWang2001,YayuWang2002,Yayu2006,Yayu2010}; while the gap at antinode gradually diverges from $d$-wave gap symmetry due to the participation of pseudogap and increases linearly with doping decreasing. That indicates the pre-forming Cooper pairs contributing to the pseudogap. For quasiparticle spectral weight, it is almost constant and cuts down a little for extremely underdoped ones at the nodal regime while monotonically decreases as doping level decreasing at anti-nodal regime. These behaviors suggest different physics are dominated in nodal and anti-nodal regions. For nodal direction specially, the band dispersion shows a ``kink'' at roughly $70$ meV \cite{Lanzara2001}. The interesting observation is that the band velocity before and after this kink evolves such differently with doping \cite{ZhongYG2019}. The lower energy dispersion before the kink has constant velocity independent with doping, and unconventionally the higher energy dispersion after the kink has a dramatically increasing velocity as doping decreasing, even surpasses its corresponding bare band velocity for the underdoped ones. 
These counter-intuitive observations show the hint of electronic fractionalization in the cuprate as indicated by the theoretical description given in this work. 
\begin{figure*}
\small
\begin{minipage}[h]{0.326\textwidth}
\boxed{
\centering
~~\includegraphics[width=\textwidth]{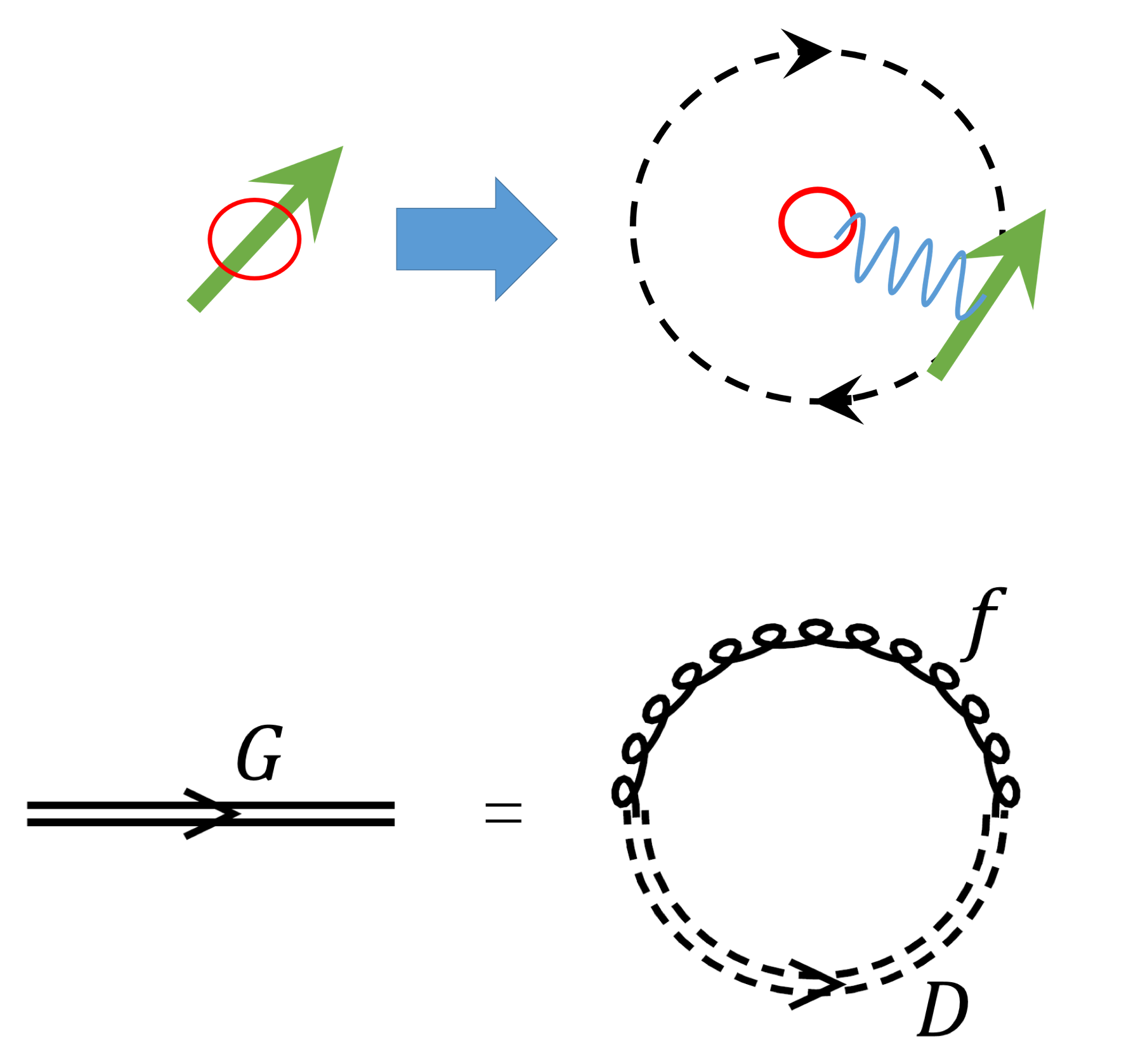}
}
\vspace{2pt}
\end{minipage}%
~~~~~~~~~~~~
\subfigure{
\begin{minipage}[p]{0.479\textwidth}
\centering
\includegraphics[width=\textwidth]{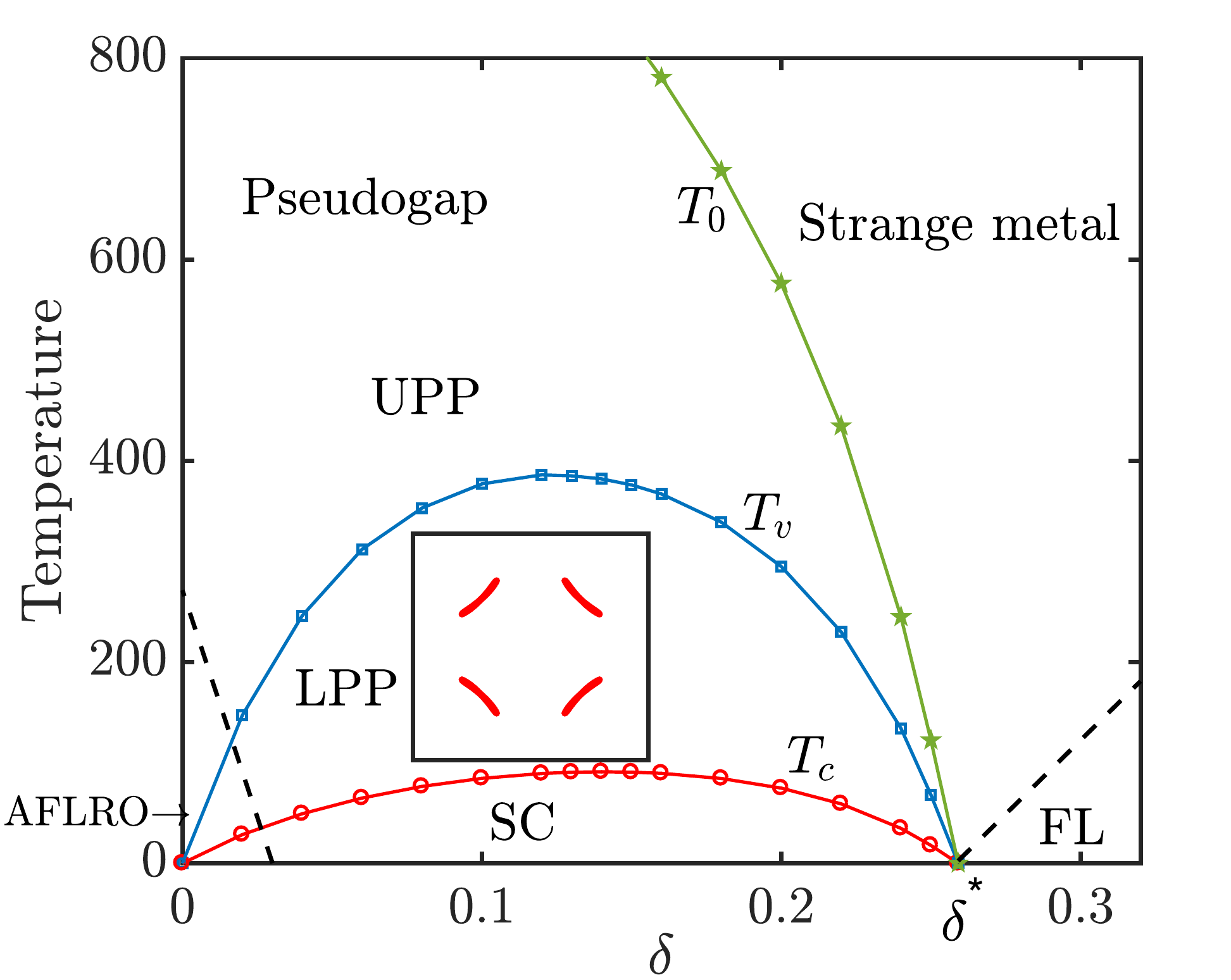}
\end{minipage}%
}
\caption{Phase-string-induced fractionalization of the electrons in the $t$-$J$ model. Left panel: the fractionalization of a doped hole; Right panel: the corresponding phase diagram in which the fractionalized particles are characterized by the mean-field state in Eq. (\ref{ground state wavefunction}) \cite{Weng2011,MaYao2014}. An exotic ``Fermi arc''  (the insert) in the pseudogap phase and the single-particle features in the superconducting state will be the main focus of the present work. Details in the left panel: A fractionalized hole is composed of a charged holon (red circle), a spinon (green arrow) and a nonlocal phase shift (blue wavy line) with internal degree of freedom shown by the dashed circle with arrows; which are described by the Feynman diagram in the bottom: the single-electron Green's function ${G}$ as a convolution of the fractionalized propagator ${D}$ and the phase factor $f$; 
Right: The fractionalized subsystems experience ODLROs to gain ``partial'' rigidity, where UPP and LPP denote the upper and lower pseudogap phases \cite{MaYao2014}, respectively, where the short-range RVB pairing is present, whereas the holons are further condensed in LPP. The true superconducting (SC) phase coherence will be realized in LPP blow $T_c$. AFLRO denotes the antiferromagnetic long-range order phase at half-filling, which may extend over a tiny but finite range of doping concentration beyond the mean field theory [dashed line], and the spin-spin correlation length is reduced with the increase of doping and eventually terminates at $\delta^*$, leading to a strange metal high-temperature behavior and a Fermi liquid (FL) instability at low temperatures at $\delta>\delta^*$.}
\label{phase diagram}
\end{figure*}

The rest of this paper is organized as follows. In Sec. \ref{section2}, we highlight some basic background of the $t$-$J$ model and its nontrivial sign structure which is called the phase string effect \cite{PhaseString1996, PhaseString1997, Weng2007, WuKai2008}. Then we briefly outline the phase-string-induced fractionalization upon hole doping and the resulting two-component RVB state, which has been obtained in the previous approach \cite {Weng2011,MaYao2014}. In Sec. \ref{section3}, based on this peculiar fractionalization and mean-field theory, we can construct a single-hole propagator by an RPA-like procedure. Such a single-particle Green's function can be regarded as a general phenomenological propagator of a single hole in the SC and pseudogap phases of a doped Mott insulator. Then in Sec. \ref{section4}, we show that in the (lower) pseudogap phase (without the SC phase coherence), the spectral function exhibits a ``Fermi arc'' phenomenon as well as the ``kink'' in the dispersion along the diagonal direction in the Brillouin zone. In particular, we show that the doping dependence of the whole features is systematically in agreement with the ARPES experiments \cite{ZhongYG2018,YGZhong2018}. Furthermore, we show that another ``kink'' or two-gap structure appears along the Fermi surface in the SC state, which replaces the Fermi arcs phenomenon in the pseudogap phase. The overall doping dependences of the two gaps, the corresponding quasiparticle spectral weights in the nodal and anitnodal regimes, respectively, as well as other spectral features, also agree with the ARPES measurements \cite{ZhongYG2018,YGZhong2018} very consistently.  Finally, Sec. \ref{section5} is devoted to summary and discussion.

\section{Background: The microscopic theory based on the $t$-$J$ model \label{section2}}

\subsection{$t$-$J$ model and the phase string effect}

The $t$-$J$ Hamiltonian $H_{t-J}=H_t+H_J$, with
\begin{align}
\small
\begin{aligned}
&H_t=-t\sum\limits_{\langle i,j\rangle,\sigma}c_{i\sigma}^\dag c_{j\sigma}+h.c.\\
&H_J=J\sum\limits_{\langle i,j\rangle}\left(\boldsymbol{S}_i\cdot\boldsymbol{S}_j-\frac{1}{4}n_in_j\right)
\end{aligned}
\label{t-J}
\end{align}
is defined on a 2D square lattice with the Hilbert space restricted by no-double-occupancy constraint:
\begin{align}
n_i\leq1
\label{no double occupancy}
\end{align}
where the electron number operator $n_i=\sum_\sigma c_{i\sigma}^\dag c_{i\sigma}$ and $\bs{S}_i$ is the corresponding $SU(2)$ spin operator.

The $t$-$J$ model can be regarded as the Hubbard model in the large-$U$ limit. The Fermi statistical sign structure is essential in determining the Landau-Fermi liquid state in the Hubbard model with sufficiently weak on-site repulsive potential $U$. But in the large-$U$ limit, the Fermion signs will get substantially \textit{suppressed} near the half-filling due to the no-double-occupancy constraint [cf. Eq. (\ref{no double occupancy})]. In particular, the Fermion signs will be totally diminished at half-filling and replaced by a much sparse sign structure away from the half-filling, which is known as the phase string \cite{PhaseString1996, PhaseString1997, Weng2007, WuKai2008,Zaanen2011,ZYWeng2011}. A continuous evolution from the fermion sign structure to the phase string sign structure with opening up the Mott gap has been also rigorously formulated for an arbitrary $U$ of the Hubbard model \cite{LongZhangHubbard}.

The no-double-occupancy constraint Eq. (\ref{no double occupancy}) and the resulting phase-string sign structure in the $t$-$J$ model imply that the model cannot be treated perturbatively in the electron representation of Eq. (\ref{t-J}). In fact, the phase-string sign structure can be mapped onto the mutual-semion statistics \cite{Weng2007, Weng2011} between the doped holes and the spins in the background, which can be accounted by a many-body nonlocal (generalized) Berry phase associated with each hole path \cite{WuKai2008}. The conventional slave-particle scheme \cite{Patrick2006} introduced to handle the constraint Eq. (\ref{no double occupancy}) should be thus generalized to properly accommodate the phase-string sign structure, which is to be outlined below as a new fractionalization scheme \cite{Weng2011} to describe such a strongly correlated system.

\subsection{Fractionalization}
Due to the aforementioned phase-string sign structure, a proper fractionalization of a bare hole created by the electron $c$-operator is given by \cite{Weng2011}:
\begin{align}
\hat{c}_{i\sigma}=h_i^\dag a_{i\bar{\sigma}}^\dag e^{i\hat{\Omega}_{i}}
\label{fractionalization}
\end{align}
where $h_i^\dag$ denotes a \emph{bosonic} holon creation operator, and $a_{i\bar\sigma}^\dag$ the creation of a \emph{fermionic} backflow spinon with the spin $\bar\sigma\equiv-\sigma$ associated with the doped hole, which in general is distributed in the spin background around the hole (cf. the left panel of Fig. \ref{phase diagram}). Equation (\ref{fractionalization}) describes a bare hole as a \emph{composite} object with internal structure, especially the  the ``phase fractionalization'' of $e^{i\hat{\Omega}_{i}}$ due to the phase string sign structure \cite{Weng2011}.  

Corresponding to such a peculiar fractionalization, which is schematically illustrated in the left panel of Fig. ~\ref{phase diagram}, a new ``mean-field'' ground state has been obtained \cite{Weng2011,MaYao2014} as follows (cf. the right panel of Fig. ~\ref{phase diagram}):
\begin{align}
|\Psi\rangle=\hat{\mathcal{P}}\Big(|\Phi_h\rangle\otimes|\Phi_a\rangle\otimes|\mathrm{RVB}\rangle\Big)
\label{ground state wavefunction}
\end{align}
in which the holons are Bose-condensed in $|\Phi_h\rangle$ and the $a$-spinons are in a BCS-like pairing state $|\Phi_a\rangle$, in addition to the bosonic RVB state $|\mathrm{RVB}\rangle$. Different from the one-component spinon in the conventional slave-particle scheme \cite{Patrick2006}, the ground state (\ref{ground state wavefunction}) has a two-component spinon structure, which is essentially required by the phase string sign structure \cite{Weng2011}.   Note that $\hat{\mathcal{P}}$ denotes a projection operator to ensure the following constraints among different fractionalized species: The holon and backflow $a$-spinon satisfy the constraint $\sum_\sigma n_{i\sigma}^a=n_i^h$ with $n_{i\sigma}^a=a_{i\sigma}^\dag a_{i\sigma}$ and $n_i^h=h_i^\dag h_i$; Furthermore, at each hole site, the spin $\bs{S}_i^a$ carried by the $a$-spinon and the spin $\bs{S}_i^b$ by the $b$-spinon must compensate each other, i.e., $\bs{S}_i^a+\bs{S}_i^b=0$ \cite{Weng2011,MaYao2014}.

The key characterization of Eq. (\ref{fractionalization}) is the phase factor $e^{i\hat{\Omega}_{i}}$, which is defined by
\begin{align}
\left\{
\begin{aligned}
&\hat{\Omega}_i\equiv\frac{1}{2}\left(\Phi_i^s-\Phi_i^0\right)\\
&\Phi_i^s=\sum\limits_{l\ne i}\theta_i(l)(n_{l\uparrow}^b-n_{l\downarrow}^b)\\
&\Phi_i^0=\sum\limits_{l\ne i}\theta_i(l)
\end{aligned}
\right.
\label{gauge field}
\end{align}
where $n_{l\sigma}^b=b_{l\sigma}^\dag b_{l\sigma}$ and $\theta_i(l)\equiv\pm\text{Im}\ln(z_i-z_l)$ [with $z_i$ ($z_l$) as the 2D complex coordinate of the site $i$ (the site $l$)]. Here $n_{l\sigma}^b$ will act on a spin background described by the Schwinger bosons, created by $b_{l\sigma}^\dag$, which form a half-filling RVB state denoted by $|\mathrm{RVB}\rangle$ \cite{Weng2011}.  $\Phi_i^s$ in Eq. (\ref{gauge field}) then represents the vortices (anti-vortices) attached to the $b$-spinons, which should be mostly compensated with each other due to the RVB pairing of the $b$-spinons in $|\mathrm{RVB}\rangle$, except for an unpaired spinon associated with hole (illustrated in the left upper panel of Fig. ~\ref{phase diagram}), after its RVB partner is removed at the hole site by $a_{i\bar\sigma}^\dag$.

Such a ground state in the single-hole limit has been recently shown to agree with the DMRG results very well by VMC method \cite{SingleHole}, where the phase string factor $e^{i\hat\Omega_i}$ reproduces the persisting circling spin current around the hole by its spin partner with the correct total angular momentum $L_z=\pm1$ as illustrated in the left panel of Fig. \ref{phase diagram}.  

At finite doping, with capturing the singular sign structure of the $t$-$J$ model via the many-body phase factor in Eq. (\ref{gauge field}), the fractionalized particles of the holon and $a$-spinon, together with the $b$-spinon in the spin background, will all behave much smoothly to be well described by a mean-field theory. At the matter of fact, three subsystems are all in off-diagonal-long-range-order (ODLRO) states in the lower pseudogap phase (LPP) and superconducting (SC) phase. Based on the mean-field calculation in Ref. \onlinecite{MaYao2014}, a general phase diagram including the antiferromagnetic long-range order (AFLRO) state (at half-filling), SC phase, LPP, upper pseudogap phase (UPP), strange metal, and a possible low-temperature Fermi liquid (in the overdoped regime), characterized by the hidden ODLROs mentioned above with transition temperatures, $T_c$, $T_v$, and $T_0$, etc., has been determined, which is summarized in the right panel of Fig. ~\ref{phase diagram}.

It is noted that even though the ground state Eq. (\ref{ground state wavefunction}) has the same hidden ODLROs (the holon condensation $\langle h_i^\dag\rangle \ne0$ and $\Delta_{ij}^0\propto\Delta_{ij}^a\ne0$ \ \cite{Weng2011,MaYao2014}) in both the LPP and SC phase, the SC phase is distinguished from the LPP in Fig.\ref{phase diagram} by having an additional \emph{true} ODLRO, i.e., the d-wave superconducting order parameter \cite{MaYao2014}:
\begin{align}\label{DSC}
\Delta_{ij}^{\text{SC}}=\Delta_{ij}^0\left\langle e^{\frac{i}{2}\left(\Phi_i^s+\Phi_j^s\right)}\right\rangle
\end{align}
with $\left\langle e^{\frac{i}{2}\left(\Phi_i^s+\Phi_j^s\right)}\right\rangle\ne0$ (which also decides the d-wave symmetry of the pairing \cite{MaYao2014}). The superconducting phase transition at $T_c$ \cite{Kivelson1995} is thus determined \cite{JWM2010} by the disassociation of vortex-antivortex binding as driven by the thermal $b$-spinon excitations in $|\mathrm{RVB}\rangle$.

The main goal of this work will be to examine the unique predictions of the phase-string-induced electron fractionalization in Eqs. (\ref{fractionalization}) and (\ref{ground state wavefunction}) that can be probed by the ARPES experiment in the LPP and SC phase. For this purpose, we shall use the same parameters in determining the phase diagram in Fig.~\ref{phase diagram} as given in Ref.  \onlinecite{MaYao2014} and further outline the underlying mean-field equations for the ground state (\ref{ground state wavefunction}) in the Appendix \ref{App.A} for the sake of being self-contained. More details can be found in Refs. \onlinecite{Weng2011,MaYao2014}.\\

\section{Single-particle Green's function\label{section3}}

\subsection{Electron fractionalization and the single-hole propagator at the mean-field level\label{Sec.fractionalization}}

The fractionalization of the doped hole in Eq. (\ref{fractionalization}) and the ground state in Eq. (\ref{ground state wavefunction}) constitute the essential microscopic description for the superconducting and lower pseudogap phases \cite{Weng2011,MaYao2014}. Such a fractionalized ground state is composed of three \textit{hidden} ODLROs in both the LPP and SC phase, i.e., the holons are always Bose-condensed, the $a$-spinons are in BCS-like pairing, and the $b$-spinons form short-range bosonic RVB pairing. In view of the holon condensation in the LPP and SC state, one may further introduce
\begin{equation}\label{twistc}
\tilde{c}_{i\sigma}\equiv h_i^\dag a_{i\bar\sigma}^\dag
\end{equation}
to denote a twisted hole as a combination of the holon and backflow spinon such that the fractionalization in Eq. (\ref{fractionalization}) is reexpressed as in Eq. (\ref{tildec}) and the ground state is rewritten compactly as in Eq. (\ref{ground state wavefunction1}). ave-like pairing amplitude in $|\Phi_a\rangle$.

\begin{figure*}
\begin{minipage}[h]{0.35\textwidth}
\centering
\includegraphics[width=\textwidth]{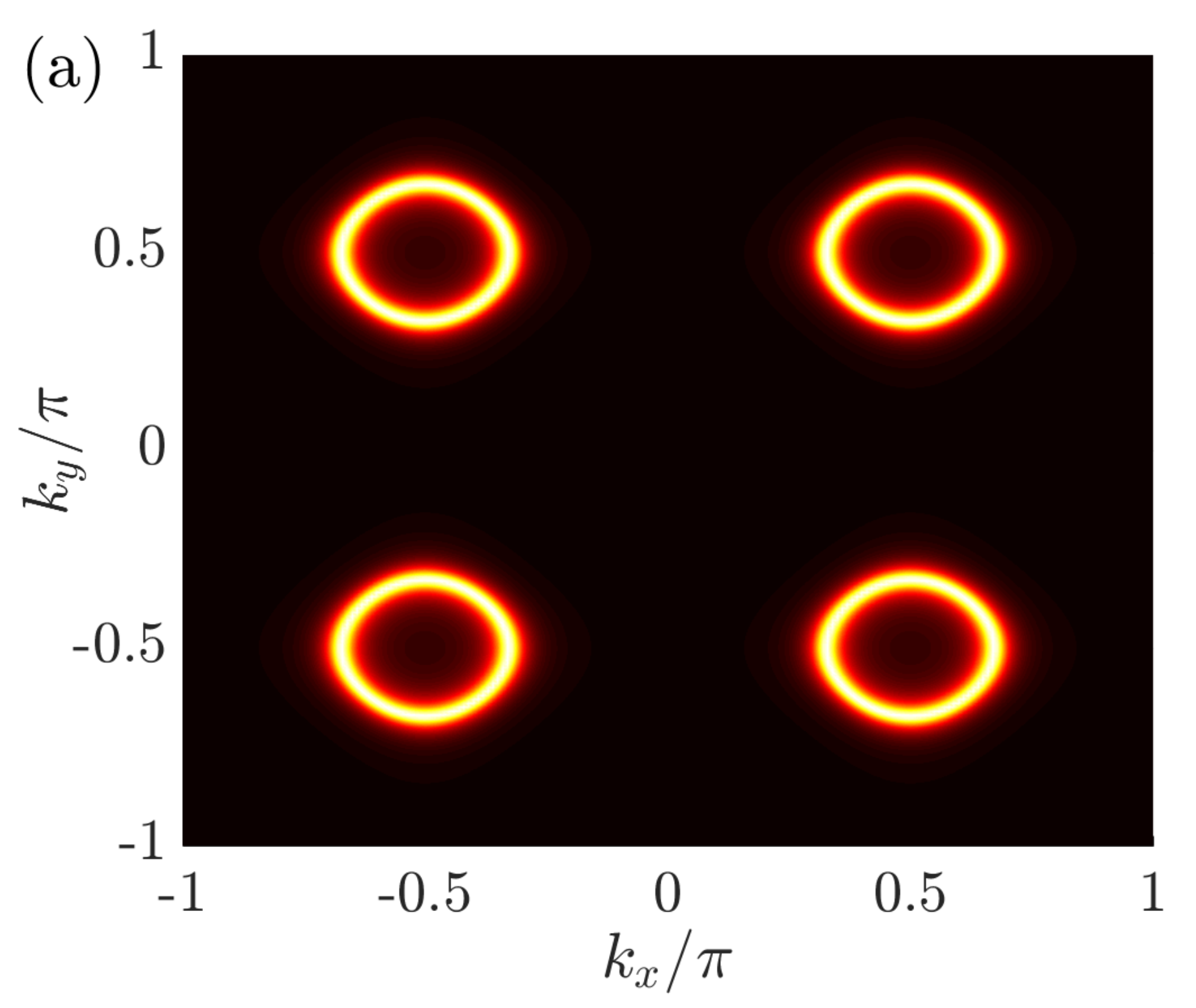}
\end{minipage}~~~~~~~~~~~
\begin{minipage}[h]{0.35\textwidth}
\centering
\includegraphics[width=\textwidth]{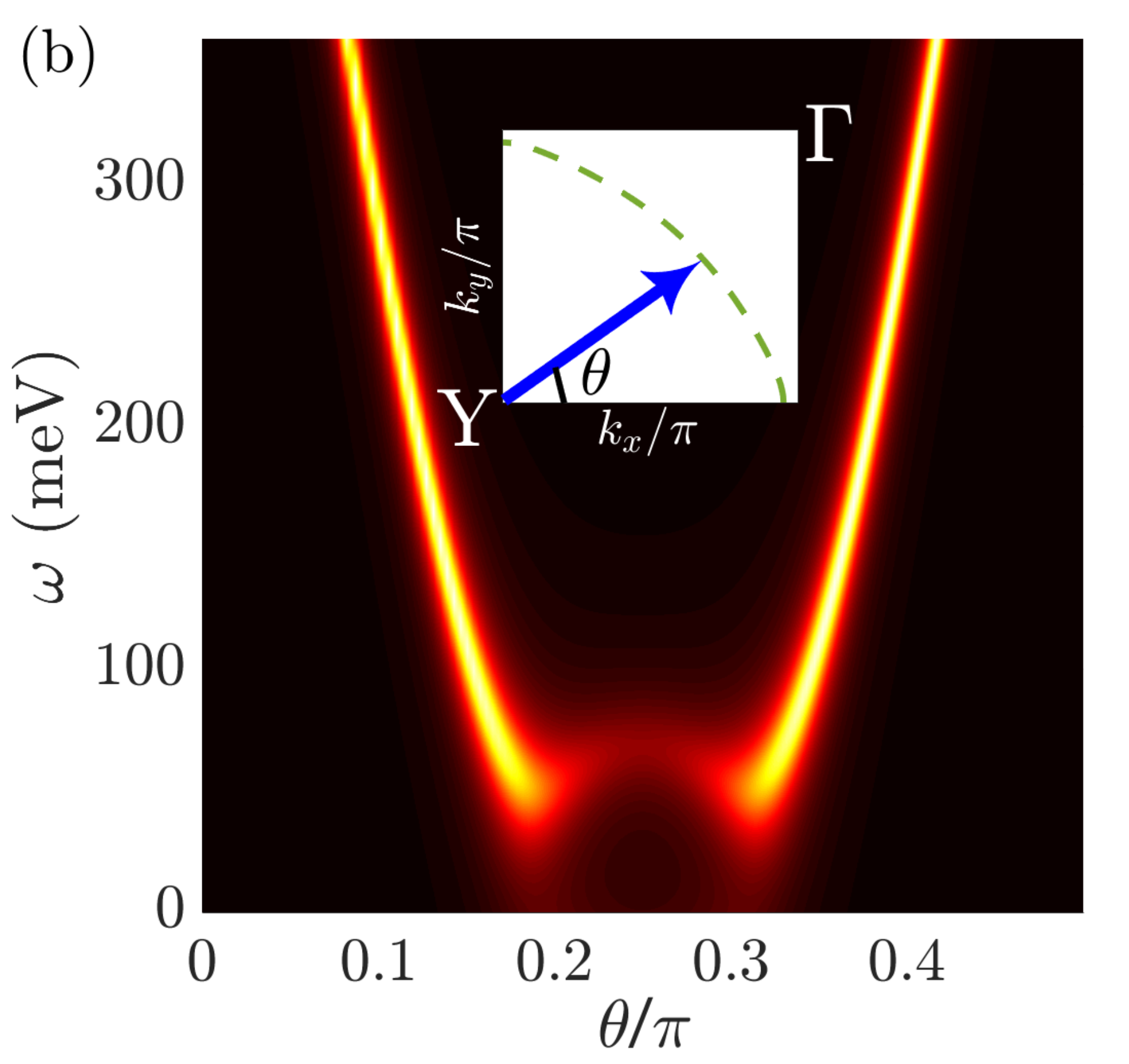}
\end{minipage}
\caption{The spectral function of the single-hole propagator at the mean-field level as given in Eq. (\ref{G11}), in which the hole is totally fractionalized according to Eq. (\ref{fractionalization}). (a) Four Fermi pockets are centered at $\bs{k}_0=(\pm\pi/2,\pm\pi/2)$; (b) An s-wave pairing for the $a$-spinons leads to a gap opening for each Fermi pocket at finite doping (cf. Appendix A). Inset: the momentum scan is shown in the Brillouin zone with $\Gamma=(0,0)$ and $Y=(\pi,\pi)$. Note that the phase fluctuation is neglected here by taking $f_0(q)$ sharply peaked at $q=0$ in Eq. (\ref{G11}). } 
\label{Fermi pocket}
\end{figure*}

Such a twisted hole, created by $\tilde{c}_{i\sigma}$, will propagate \textit{coherently} in the LPP and SC phase. Denoting its propagator as $\hat{D}(i,j;\tau)$ between sites $i$ and $j$ and imaginary time $\tau$. Its leading term (mean-field) $\hat{D}_0(i,j;\tau)$ is then given by:
\begin{align}
\hat{D}_0(i,j;\tau)=D^h_0\hat{D}_0^a(i,j;\tau),
\label{fractionalized}
\end{align}
where the condensed holon propagator $D_0^h\sim\delta$ and $\hat{D}_0^a$ depicts the propagator of the $a$-spinons, which are in the $s$-wave BCS-like state with a $2\times2$ Nambu-Gor'kov propagator given by \cite{MaYao2014}
\begin{align}
\renewcommand\arraystretch{1.5}
\hat{D}_0^a(\bs{k},\omega)=\left(
\begin{array}{ccc}
G^a_{\uparrow\uparrow}(\bs{k},\omega) & F^a(\bs{k},\omega)\\
F^a(\bs{k},\omega) & -G^a_{\downarrow\downarrow}(-\bs{k},-\omega)~.
\end{array}
\right)
\end{align}
Here $G^a(\bs{k},\omega)$/$F^a(\bs{k},\omega)$ are the normal/anomalous components of the standard BCS-type Green's function: 
\begin{align}
\left\{
\begin{aligned}
&G^a_{\uparrow\uparrow}(\boldsymbol{k},\omega)=\frac{u_k^2}{\omega-\epsilon_{k1}^a}+\frac{v_k^2}{\omega+\epsilon_{k1}^a}=G^a_{\downarrow\downarrow}(\boldsymbol{k},\omega)\\
&F^a(\boldsymbol{k},\omega)=u_kv_k\left(\frac{1}{\omega-\epsilon_{k1}^a}-\frac{1}{\omega+\epsilon_{k1}^a}\right)
\end{aligned}
\right.
\label{Nambu-Gor'kov}
\end{align}
with the coefficients determined by the mean-field theory \cite{MaYao2014} [cf. Eq. (\ref{mean-field}) in Appendix \ref{App.A}]. 
\begin{align}
\left\{
\begin{aligned}
&u_k^2=\frac{1}{2}\left(1+\frac{\xi_{k1}^a}{\epsilon_{k1}^a}\right)\\
&v_k^2=\frac{1}{2}\left(1-\frac{\xi_{k1}^a}{\epsilon_{k1}^a}\right)
\end{aligned}
\right.
\end{align}
where $\xi_{k1}^a=-2\tilde{t}_a\sqrt{\cos^2(k_xa_0)+\cos^2(k_ya_0)}+\lambda_a,~\tilde{t}_a=t_a+\gamma\chi^a$, is the energy spectrum for the $a$-spinon gas before pairing and $\epsilon_{k1}^a=\sqrt{(\xi_{k1}^a)^2+(\Delta_k^a)^2}$ is the Bogoliubov energy spectrum with a gap $\Delta_k^a=2\gamma\Delta^a\sqrt{\cos^2(k_xa_0)+\cos^2(k_ya_0)}$, which shows a $k$-dependence in the presence of the $\pi$-flux in $H_a$ (Appendix A) even though the pairing order parameter $\Delta^a$ is s-wave-like \cite{MaYao2014}. 

Then according to Eq.  (\ref{tildec}), the single-hole Green's function can be expressed by
\begin{align}\label{G}
[\hat{G}(i,j;\tau)]_{11}&=[\hat{D}_0(i,j;\tau)]_{11}\left\langle e^{i\left[\hat{\Omega}_i(\tau)-\hat{\Omega}_j(0)\right]}\right\rangle\nonumber\\
&= [\hat{D}_0(i,j;\tau)]_{11} f(i,j;\tau)~, \\
[\hat{G}(i,j;\tau)]_{12}&=[\hat{D}_0(i,j;\tau)]_{12}\left\langle e^{i\left[\hat{\Omega}_i(\tau)+\hat{\Omega}_j(0)\right]}\right\rangle\nonumber\\
& \simeq [\hat{D}_0(i,j;\tau)]_{12} \left\langle e^{2i\hat{\Omega}_j(0)}\right\rangle f(i,j;\tau)~,
\end{align}
where
\begin{equation}\label{ff}
f(i,j;\tau)\equiv \left \langle e^{i\left[\hat{\Omega}_i(\tau)-\hat{\Omega}_j(0)\right]}\right\rangle~.
\end{equation}
Note that in the anomalous term $[\hat{G}]_{12}$, $\left\langle e^{2i\left[\hat{\Omega}_j(0)\right]}\right\rangle\neq 0$ in the SC state and vanishes in the LPP. It is consistent with the phase coherence of the phase factor $e^{i\hat{\Omega}_i}$ controlling the SC order parameter in Eq. (\ref{DSC}).

It is interesting to examine the phase structure of $e^{i\hat{\Omega}_i}$ which renormalizes the single-particle propagator via $f(i,j;\tau)$ in Eq. (\ref{ff}). One may rewrite Eq. (\ref{ff}) in terms of Eq. (\ref{gauge field}) as follows
\begin{align}
f(i,j;\tau)=&\left\langle \exp\left\{-\frac{i}{2}\left[\Phi_i^s(\tau)-\Phi_j^s(0)\right]\right\}\right\rangle\nonumber\\
&\cdot\exp{\left\{\frac{i}{2}\left(\Phi_i^0-\Phi_j^0\right)\right\}}~,
\label{phase correlation function}
\end{align}
by which one may further express
\begin{align}
&\exp{\left\{\frac{i}{2}\left(\Phi_i^0-\Phi_j^0\right)\right\}}\nonumber\\
&=\exp{\left\{\frac{i}{2}\left(\Phi_i^0-\Phi_{i_1}^0+\Phi_{i_1}^0-\Phi_{i_2}^0+....+\Phi_{i_n}^0-\Phi_{j}^0\right)\right\}}\nonumber \\
&=\exp{\left\{{i}\left(\phi_{ii_1}^0+\phi_{i_1i_2}^0+....+\phi_{i_nj}^0\right)\right\}}\nonumber \\
&\cdot \prod_{i\rightarrow j}\exp\left\{\frac{i}{2}\left[\theta_{i_s}(i_{s+1})-\theta_{i_{s+1}}(i_s)\right]\right\}~,
\end{align}
in which one inserts a sequences of the nearest-neighboring links connecting $i$ and $j$: $i_1$, ${i_2}$, ...., $i_n$. By noting 
\begin{align}
&\prod_{i\rightarrow j}\exp\left\{\frac{i}{2}\left[\theta_{i_s}(i_{s+1})-\theta_{i_{s+1}}(i_s)\right]\right\}\nonumber\\
=&\left(e^{\pm i\frac{\pi}{2}}\right)^{i-j}
\end{align}
with using $\theta_{i_s}(i_{s+1})-\theta_{i_{s+1}}(i_s)=\pm \pi$, one finally arrives at
\begin{align}
f(i,j;\tau)&=e^{i\bs{k}_0\cdot(\bs{r}_i-\bs{r}_j)} f_0(i,j;\tau)~,
\label{f}
\end{align}
with
\begin{align}
f_0(i,j;\tau)\equiv  \left\langle \exp\left\{-\frac{i}{2}\left[\Phi_i^s(\tau)-\Phi_j^s(0)\right]\right\}\right\rangle~
\label{f0}
\end{align}
and
\begin{equation}
\boldsymbol{k}_0\equiv \left(\pm\frac{\pi}{2},\pm\frac{\pi}{2}\right)~.
\label{k0}
\end{equation}
Note that in obtaining Eq. (\ref{f}), there is an additional phase factor $\exp{\left\{{i}\sum_{i\rightarrow j} \phi_{i_si_{s+1}}^0\right\}}$, which should be combined with $\hat{D}_a^0$ to make it gauge invariant (noting that the $a$-spinon sees $\phi_{i_si_{s+1}}^0$ in the Hamiltonian $H_a$ [Eq. (\ref{mean-field})]. By taking a special gauge of the $\pi$ flux, $\exp{\left\{{i}\sum_{i\rightarrow j} \phi_{i_si_{s+1}}^0\right\}}=(-1)^{i_y-j_y}$ or $(-1)^{i_x-j_x}$, which can be simply absorbed into the oscillating phase factor $e^{i\bs{k}_0\cdot(\bs{r}_i-\bs{r}_j)}$ to connect the four finite momenta given in Eq. (\ref{k0}). 
 
Finally, the single-hole propagator may be approximately reduced to the following form 
\begin{align}\label{G11}
\hat{G}(\boldsymbol{k},\omega)&\simeq \sum_q f_0(q) \hat{D}_0(\boldsymbol{k}-\boldsymbol{k}_0-\boldsymbol{q},\omega-q_0)~
\end{align}
with $q=( \boldsymbol{q},q_0)$. In particular, in the weak phase fluctuation case, like in the SC phase, $f_0(q)$ is sharply peaked at $q=0$, where the spectral function corresponding to Eq. (\ref{G11}) will exhibit four Fermi pockets as shown in Fig. ~\ref{Fermi pocket}(a). Note that
the leading phase factor of $\bs{k}_0$ in Eq. (\ref{f}) will shift the Fermi pockets of the $a$-spinon centered at momentum ($0$,$0$) and ($\pm\pi$,$0$) or  ($0$, $\pm\pi$) (depending on the gauge choice of $\phi_{i_si_{s+1}}^0$)  \cite{MaYao2014} to four $a$-spinon Fermi pockets (each with a Luttinger volume of $\delta/4$) centered at the momenta $\boldsymbol{k}_0$ given in Eq. (\ref{k0}). Figure ~\ref{Fermi pocket}(b) further indicates a gap opened up in each of the Fermi pockets as the $a$-spinons are in the s-wave BCS pairing \cite{MaYao2014} [cf. Appendix \ref{App.A}].

Therefore, an injected hole created by ARPES will in general be fractionalized according to the scheme of Eq. (\ref{fractionalization}). To the leading order of approximation, the single-particle propagator is described in Eq. (\ref{G11}) by four s-wave gapped Fermi pockets locating at $\boldsymbol{k}_0$ along the magnetic Brillouin zone boundary. There is no trace of a large Fermi surface of the electrons at this level. However, in the following, we shall go beyond the mean-field level to show that the twisted hole propagator $\hat{D}$ will have a Landau quasiparticle or Bogoliubov quasiparticle excitation emerging within the gap as an RPA-like correction, which will then substitute $\hat{D_0}$ in Eq. (\ref{G11}) to get the full single-particle Green's function in the LPP and SC phase.

\subsection{Quasiparticle as an emergent mode\label{Sec.QP}}

Note that in the ground state (\ref{ground state wavefunction}) or (\ref{ground state wavefunction1}), there is no trace of the electrons, which are all fractionalized according to Eqs. (\ref{fractionalization}) or (\ref{tildec}). Hence, if a gapless quasiparicle excitation exists, it must be considered as an emergent ``collective'' mode as a bound state of the fractionalized particles. Physically, the residual interaction in the $t$-$J$ model beyond the mean-field theory should provide the intrinsic binding force to realize such a ``collective'' excitation, if it exits. In the earlier slave-boson fractionalization scheme, such a quasiparticle as a bound state has been studied by adding an attractive potential artificially \cite{Ng2005}. 

Alternatively, an equation-of-motion approach has been proposed in the phase string formulation of the $t$-$J$ model, in which a quasiparticle as a stable entity over a finite-time scale, before its decay into fractional particles in a long time, can be described beyond the mean-field theory without adding an artificial attractive potential. Namely, if a bare hole is created in the ground state (\ref{ground state wavefunction}) by $\hat{c}_{i\sigma}$, it will evolve with the time as follows \cite{Weng2011,MaYao2014}:
\begin{eqnarray}\label{eom}
& & \ \ \ \ -i\partial_t \hat{c}_{i\sigma}|\Psi_G\rangle =  [H_{t-J},\hat{c}_{i\sigma}]|\Psi_G\rangle \nonumber \\
&=& \left[t_{\mathrm{eff}}\sum\limits_{j=NN(i)} \hat{c}_{j\sigma} +\mu c_{i\sigma}  -J\sum\limits_{j=NN(i)} \Delta_{ij}^{\mathrm{SC}} \sigma \hat{c}_{j\bar\sigma}^\dagger \right]|\Psi_G\rangle \nonumber \\
                           &+& \text{scattering term}+ \text{decay term},
\end{eqnarray}
which shows that the bare hole will first coherently propagate in a (Bogoliubov) single-particle fashion with the SC order parameter $\Delta_{ij}^{SC}$ before its decay into fractionalized particles. In fact, previously this propagation of the hole has been treated as a renormalized mean-field solution based on the $t$-$J$ model as follows.

Here, without considering the scattering and decay processes, the bare-hole does follow a renormalized band-structure quasiparticle behavior in Eq. (\ref{eom}), which is described by $\hat{G}_0$ in a form of the standard Nambu-Gor'kov Green's function of a $d$-wave BCS pairing state as \cite{MaYao2014}: 
\begin{align}
\renewcommand\arraystretch{1.5}
\hat{G}_0(\boldsymbol{k},\omega)=\left(
\begin{array}{ccc}
G^{0}_{\uparrow\uparrow}(\boldsymbol{k},\omega) & F^c_0(\boldsymbol{k},\omega)\\
F^c_0(\boldsymbol{k},\omega) & -G^{0}_{\downarrow\downarrow}(-\boldsymbol{k},-\omega)
\end{array}
\right)
\label{G0}
\end{align}
where
\begin{align}
\left\{
\begin{aligned}
&G^0_{\sigma\sigma}(\boldsymbol{k},\omega)=\frac{u_k^2}{\omega-E_k}+\frac{v_k^2}{\omega+E_k}~~\sigma=\uparrow,\downarrow\\
&F_0(\boldsymbol{k},\omega)=u_kv_k\left(\frac{1}{\omega-E_k}-\frac{1}{\omega+E_k}\right)
\end{aligned}
\right.
\end{align}
with the parameters given by
\begin{align}
\left\{
\begin{aligned}
&u_k^2=\frac{1}{2}\left(1+\frac{\epsilon_k^0}{E_k}\right)\\
&v_k^2=\frac{1}{2}\left(1-\frac{\epsilon_k^0}{E_k}\right)
\end{aligned}
\right.,~~E_k=\sqrt{(\epsilon_k^0)^2+\left[\Delta^{\mathrm{SC}}(\bs{k})\right]^2}
\end{align}
Here the $d$-wave gap function takes the simplest form \cite{MaYao2014}:
\begin{align}
\Delta^{\mathrm{SC}}(\bs{k})=J_{\mathrm{eff}}\Delta^a[\cos(k_xa_0)-\cos(k_ya_0)]
\label{SCgap}
\end{align}
where $\cos(k_xa_0)-\cos(k_ya_0)$ is the $d$-wave factor for the nearest-neighbor pairing. In particular, in the band dispersion
\begin{align}
\epsilon_k^0=&-2t_{\mathrm{eff}}\left[\cos(k_xa_0)+\cos(k_ya_0)\right]\nonumber\\
&-4t'\cos(k_xa_0)\cos(k_ya_0)+\mu~,
\end{align}
the the second nearest hopping $t'=-0.3t_{\mathrm{eff}}$ has been added to the $t$-$J$ model in order to compare with the ARPES experiment. 

Note that the scattering term in Eq. (\ref{eom}) is given by \cite{Weng2011,MaYao2014}
\begin{eqnarray}
&&\mathrm{scattering}\text{ \textrm{term}}=\sum_{j=NN(i)}\left[ t\left(
c_{j\sigma }\sigma S_{i}^{bz}+c_{j-\sigma }S_{i}^{b-\sigma }\right) \right.\nonumber\\
&&\left.-\frac{J}{2}\left( c_{i\sigma }\sigma S_{j}^{bz}+c_{i-\sigma }S_{j}^{b-\sigma
}\right) \right] |\Psi _{\mathrm{G}}\rangle  \label{scatt}
\end{eqnarray}
which shall be omitted for simplicity, since it involves the scattering between the quasiparticle and the background spin AF excitations in terms of the $b$-spinons which are all gapped \cite{Weng2011,MaYao2014}. Its contribution to the higher energy quasiparticle excitation can be further considered in a future study [its contribution to the SC pairing has been already incorporated into the coherent term in Eq. (\ref{eom})]. 

However, the $\text{decay term}$ \cite{Weng2011} in Eq. (\ref{eom}) indicates that the doped hole will be eventually fractionalized, due to the strong scattering of the doped hole with the background under the no-double-occupancy constraint that cannot be properly described by conventional self-energy of a bare hole (electron) interacting with a bosonic mode. Instead, its leading term is represented by fractionalizing into $ \tilde{c}_{j\sigma} $, $ \tilde{c}^{\dagger}_{j\bar{\sigma}} $,  and  $e^{i\hat{\Omega}_i}$ of the form \cite{Weng2011,MaYao2014}:
\begin{eqnarray}\label{decay}
&&\text{decay term} \\
&\propto& \left(\sum_{j=NN(i)} \tilde{c}_{j\sigma} e^{i\hat{\Omega}_i} + \sum_{j=NN(i)} \tilde{c}^{\dagger}_{j\sigma} e^{i\hat{\Omega}_i} + ...\right)|\Psi_G\rangle \nonumber
\end{eqnarray}
which will then evolve independently according to the mean-field Hamiltonians in Appendix \ref{App.A}. Here the overall amplitude of the decay term in Eq. (\ref{decay}) can be estimated $|\lambda| \sim \delta J$ \cite{Weng2011,MaYao2014}. As an inverse process of this decay, the twisted hole $ \tilde{c}$ and the phase factor $e^{i\hat{\Omega}}$ can also be recombined into a quasiparticle as given in Eq. (\ref{tildec}) with the same amplitude $\sim \lambda$. In the following, we shall incorporate such a reemerging quasiparticle component into the bare (mean-field) propagator of $\tilde{c}$, perturbatively, in terms of the small coupling strength $\lambda$.

\subsection{The single-hole Green's function: Two-component structure}

In Sec. \ref{Sec.fractionalization}, we have discussed that the quasiparticle is totally fractionalized to a twisted hole $\tilde{c}$ according to Eq. (\ref{tildec}), which is described by the propagator $\hat{D}_0$ in Eq. (\ref{fractionalized}) at the mean-field level. Then in Sec. \ref{Sec.QP}, we have shown that the quasiparticle may still recover its partial coherent motion over some finite scales of length and time according to Eq. (\ref{eom}) beyond the mean-field fractionalization. It implies that the twisted hole may also ``decay'' back into a quasiparticle,  i.e., $\tilde{c}_{i\sigma}\rightarrow \hat{c}_{i\sigma}e^{-i\hat{\Omega}_i}$ as a higher order process beyond the mean-field approximation.  

\begin{figure}[t]
\centering\includegraphics[width=0.49\textwidth]{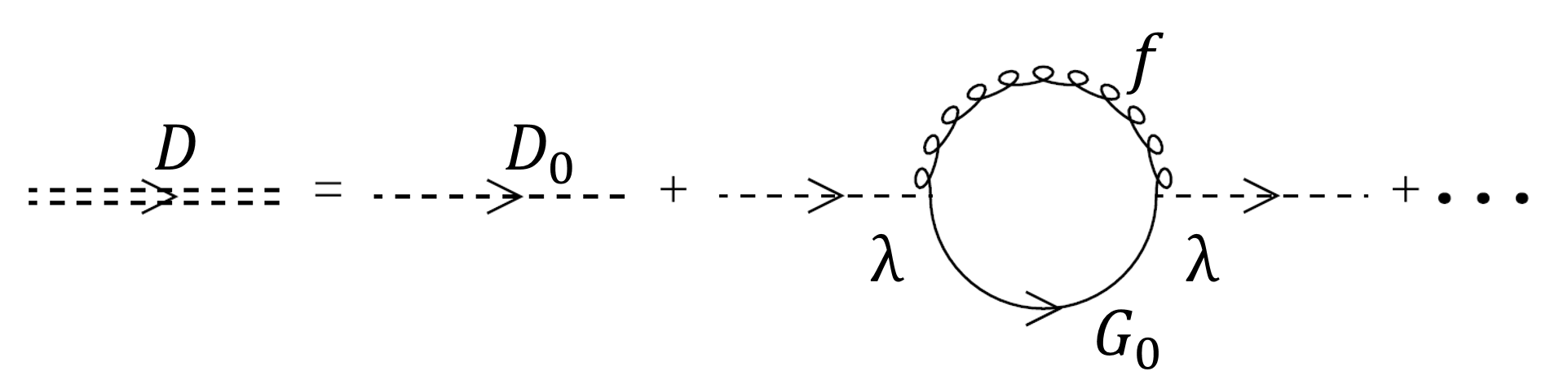}
\caption{The schematic Feynman diagram of an ``RPA'' like procedure for the twisted-hole Green's function $D$ given in Eq. (\ref{D0}), and the resulting compact form in Eq. (\ref{D}) characterizes the fractionalization by a two-component structure composed of the leading fractionalized propagator $D_0$ and the recombined bare-hole propagator $G_0$ modulated by the phase string phase factor $f$ with a vertex coupling constant ${\lambda}$.}
\label{Feynman}
\end{figure}

Now we treat the mean-field $\hat{D}_0$ as the leading term and consider the RPA-like correction to the propagator of the $\tilde{c}$ beyond $\hat{D}_0$. At an RPA level involving the decay and recombination process [cf. Eq. (\ref{decay})] as a perturbation expansion in terms of $\lambda$, the full propagator of the twisted quasiparticle, illustrated by the Feynman diagram in Fig. \ref{Feynman}, may be expressed by the following Dyson equation

\begin{align}\label{D0}
&\hat{D}(i,j;\tau)=\hat{D}_0(i,j;\tau)+\int\int d\tau'd\tau''\sum_{j',j"}\nonumber\\
&\hat{D}_0(i,j';\tau')|\lambda|^2\tilde{G}_0(j',j";\tau"-\tau')\hat{D}_0(j",j;\tau-\tau")+...
\end{align}
or in the momentum-frequency space
\begin{align}
\hat{D}(\boldsymbol{k},\omega)&=\hat{D}_0(\bs{k},\omega)+\hat{D}_0(\bs{k},\omega)|\lambda|^2\tilde{G}_0(\bs{k},\omega)\hat{D}_0(\bs{k},\omega)+...\nonumber\\
&=\frac{1}{\hat{D}_0^{-1}(\boldsymbol{k},\omega)-|\lambda|^2 \tilde{G}_0(\boldsymbol{k},\omega)}~.
\label{D}
\end{align}
where the momentum dependence of the coupling strength $\lambda$ has been omitted for simplicity.

\begin{figure*}
\includegraphics[width=0.99\textwidth]{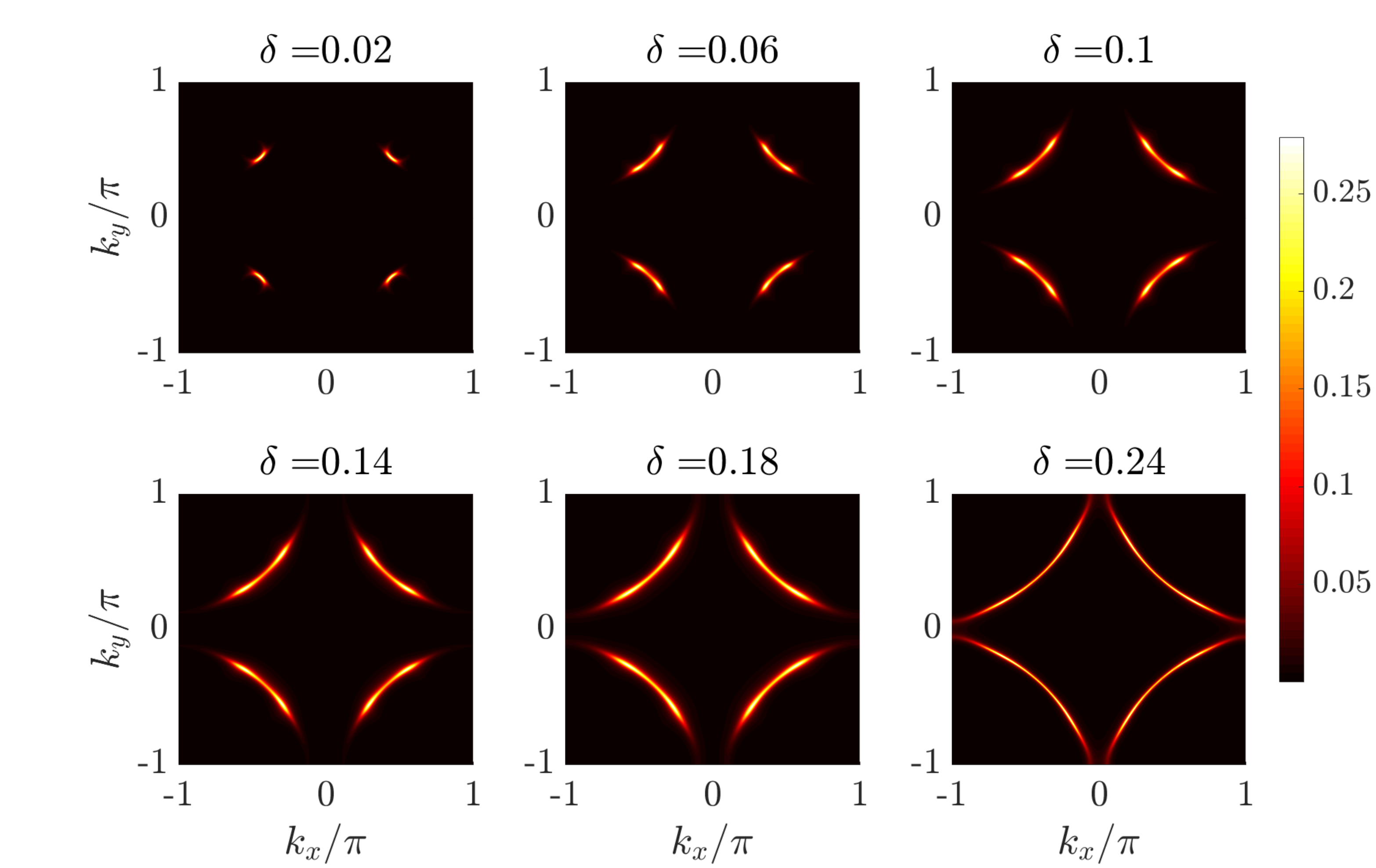}
\caption{A systematic evolution of the Fermi arcs at various doping concentrations in the LPP. The Fermi arcs will gradually evolve into a full Fermi surface in the overdoped regime, while they collapse and jump to four points at $(\pm\pi/2,\pm\pi/2)$ in the half-filling limit (see the text).} 
\label{arc}
\end{figure*}

Here in Eq. (\ref{D}), $\tilde{G}_0(\bs{k},\omega)$ is the Fourier transformation of the following convolution of the propagators of the quasiparticle $\hat{G}_0$ in Eq. (\ref{G0}) and the phase factor $e^{-i\hat{\Omega}_{i}}$, e.g., 
\begin{align}\label{tildeG011}
[\tilde{G}_0(i,j;\tau)]_{11}&=G^{0}_{\uparrow\uparrow}(i,j;\tau)f^*(i,j;\tau)~,\\
[\tilde{G}_0(i,j;\tau)]_{12}&\simeq F^c_0(i,j;\tau) \left\langle e^{2i\hat{\Omega}_j(0)}\right\rangle f^*(i,j;\tau)~.
\end{align}
By noting $\left\langle e^{2i\hat{\Omega}_j(0)}\right\rangle\neq 0$ in the SC state and $=0$ in the LPP, one may further rewrite
\begin{equation}\label{G01}
[\tilde{G}_0(\boldsymbol{k},\omega)]_{11(2)}=\sum_{{q}}f_0(q) [\hat{G}_0(\boldsymbol{k}+\boldsymbol{k}_0+\boldsymbol{q},\omega+q_0) ]_{11(2)}
\end{equation} 
by using Eq. (\ref{f}),  with $\hat{G}_0(\boldsymbol{k},\omega) $ given in Eq. (\ref{G0}) in which the anomalous term $F^c_0$ renormalized by $\left\langle e^{2i\hat{\Omega}_j(0)}\right\rangle$. 

Then, with the twisted hole propagator $\hat{D}$ obtained in Eq. (\ref{D}) to replace $\hat{D}_0$ in Eq. (\ref{G11}), the full single-particle Green's function is now given by
\begin{align}\label{G2}
\hat{G}(\boldsymbol{k},\omega)&=\sum_q f_0(q)[\hat{D}(\boldsymbol{k}-\boldsymbol{k}_0-\boldsymbol{q},\omega-q_0)~.
\end{align}
Based on Eq. (\ref{G2}), the spectral function $A(\boldsymbol{k},\omega)$ can be determined, which is directly connected to the ARPES measurement as to be discussed in the next section.

Finally, one may further simplify the full single-hole propagator to the following compact form
\begin{align}\label{G3}
\hat{G}(\boldsymbol{k},\omega) =\frac{1}{\hat{\cal D}_0^{-1}(\boldsymbol{k},\omega)-|{\lambda}|^2 \hat{G}_0(\boldsymbol{k},\omega)}~,
\end{align}
by noting that in the SC phase, due to the phase coherence of $\left\langle e^{i\hat{\Omega}_i}\right\rangle\neq 0$, the leading $f_0({q}) $ may be taken as a $\delta$-function with $q$ peaked at $q=0$, such that 
\begin{equation}\label{G02}
\tilde{G}_0(\boldsymbol{k},\omega)\simeq F_0 \hat{G}_0(\boldsymbol{k}+\boldsymbol{k}_0,\omega),
\end{equation} 
where $F_0=\sum_qf_0(q)$, and $\hat{\cal D}_0(\boldsymbol{k},\omega)=F_0^{-1}\hat{D}_0^{-1}(\boldsymbol{k}-\boldsymbol{k}_0,\omega)$. In the LPP, if one assumes that the characteristic scale of $f_0(i,j;\tau)$ is still much larger than the bare-hole propagator $\hat{G}_0$ as the decay is weak, the similar expression still holds. With the increase of the fluctuation in $f_0$, to the leading order, one may modify the high-energy component $\hat{\cal D}_0$ in Eq. (\ref{G3}) by
\begin{align}\label{G4}
\hat{\cal D}_0(\boldsymbol{k},\omega)&\equiv\sum_q f_0(q) \hat{D}_0(\boldsymbol{k}-\boldsymbol{k}_0-\boldsymbol{q},\omega-q_0)~.
\end{align}

Therefore, Eq. (\ref{G3}) is explicitly composed of two components. One is the fractionalized Green's function $\hat{\cal D}_0(\boldsymbol{k},\omega)$ given in Eq. (\ref{G4}), which essentially describes four Fermi pockets centered at $\boldsymbol{k}_0$ that are gapped by an s-wave pairing gap in the LPP and SC phase and modulated by the phase fluctuation via $f_0$. The other is the bare quasiparticle $\hat{G}_0$ with a large Fermi surface, which is incorporated in Eq. (\ref{G3}) in an RPA fashion. The interplay between these two components will determine the general structure of the spectral function $A(\boldsymbol{k},\omega)$ as to be shown in the next section. Thus, the ARPES experiment can provide a direct probe into the composite structure of strongly correlated electrons which lies in the basic mathematical description of a doped Mott insulator system. 
  
\begin{figure*}
\begin{minipage}[h]{0.32\textwidth}
\centering
\includegraphics[width=\textwidth]{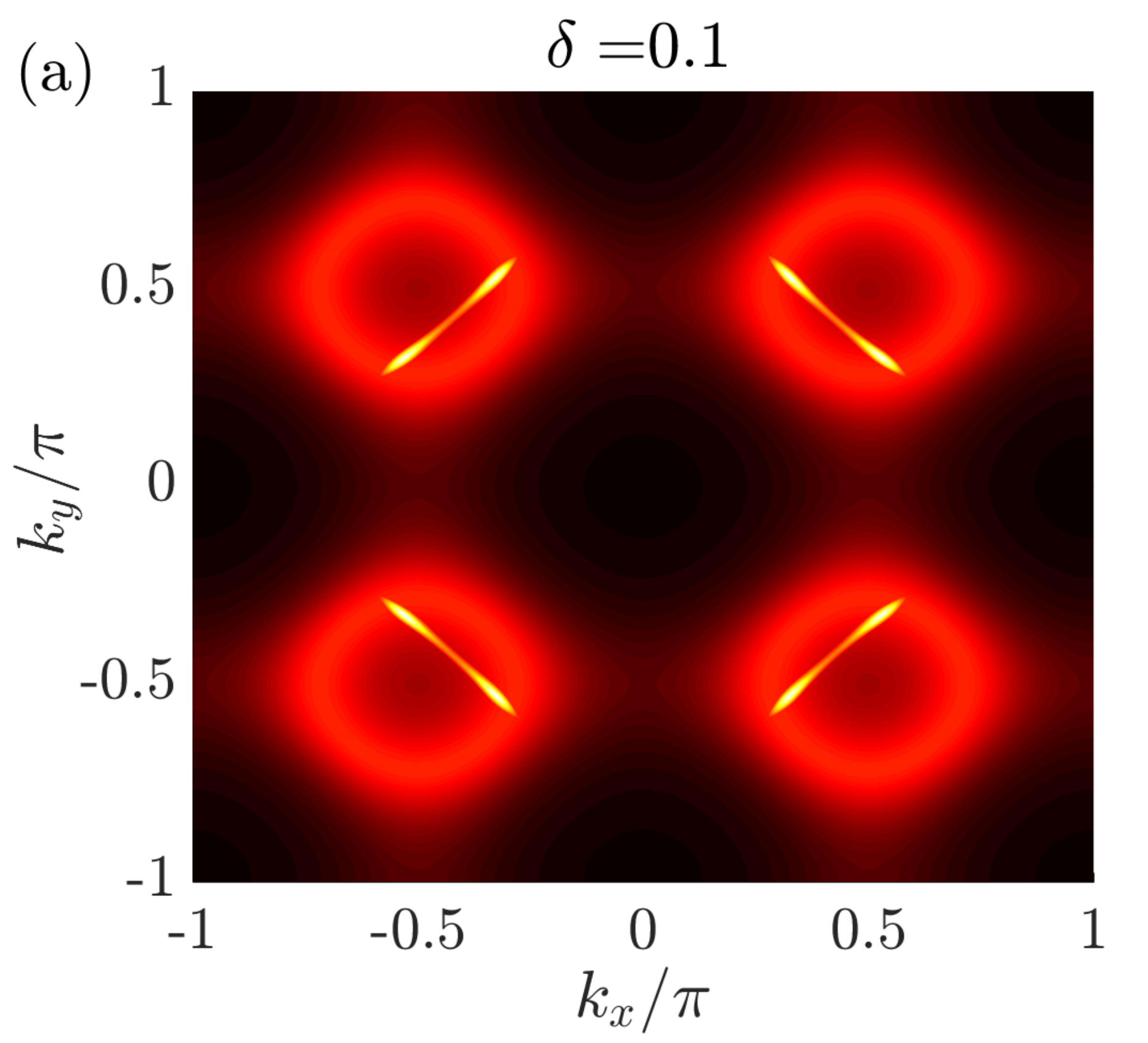}
\end{minipage}
\begin{minipage}[h]{0.29\textwidth}
\centering
\includegraphics[width=\textwidth]{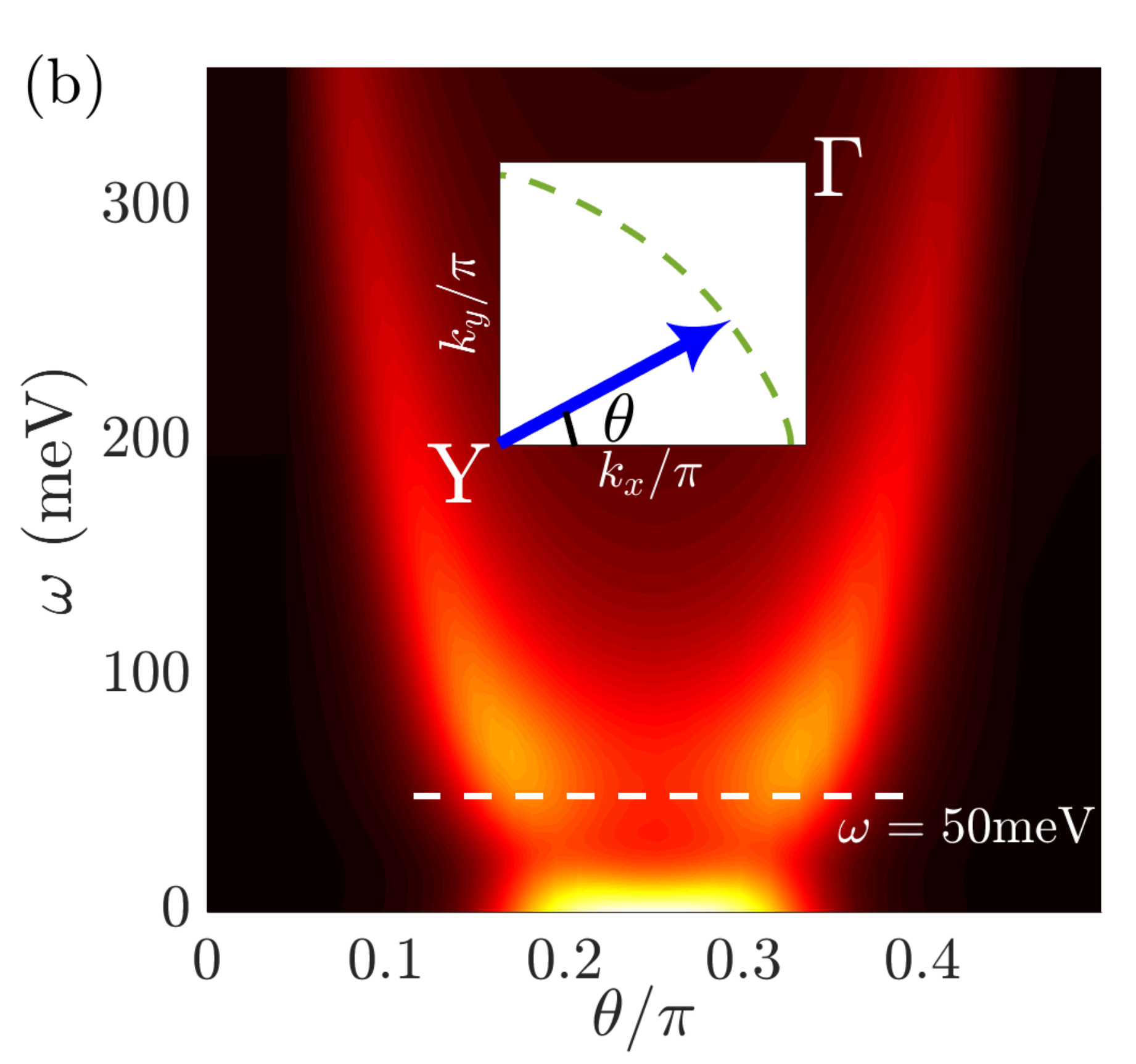}
\end{minipage}
\begin{minipage}[h]{0.37\textwidth}
\centering
\includegraphics[width=\textwidth]{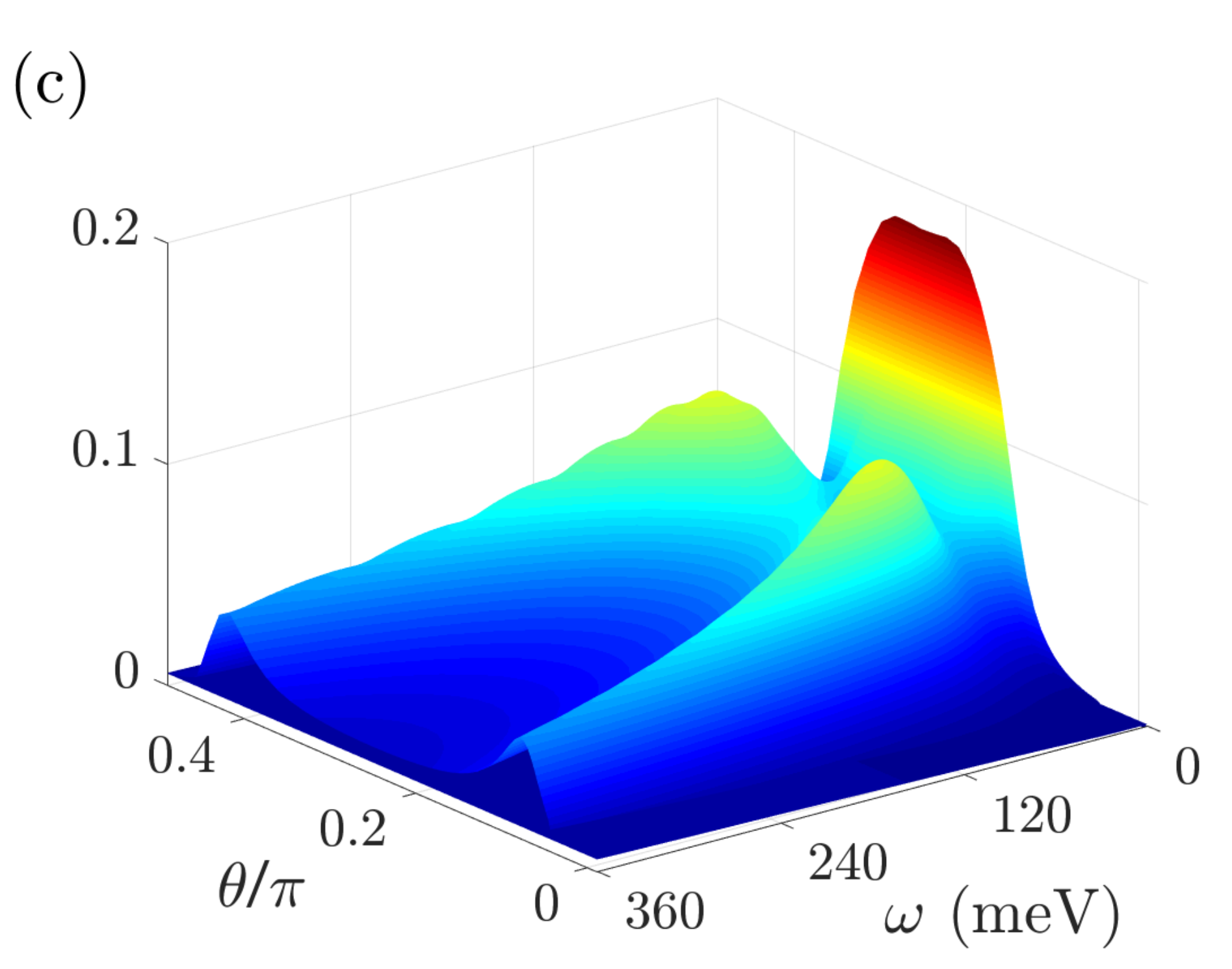}
\end{minipage}
\caption{Two-component structure of the full spectral function $A(\bs{k},\omega)$ in the LPP is shown in (a): The ${\bf k}$-space projection indicates both Fermi arcs and Fermi pockets, which are at $\omega=0$ and $\omega=50$ meV, respectively;  The $\omega$-dependence of the spectral function around one Fermi pocket is shown in (b), where the higher-$\omega$ branch corresponds to the contribution mainly from $\hat{\cal D}_0$, while the lower branch at $\omega\sim 0$ is from $\hat{G}_0$. Here $\bs{k}$ scans along the bare large Fermi surface denoted by the green dashed curve in the insert, specified by an angle $\theta$; (c) A three-dimensional illustration of $A(\bs{k},\omega)$ in (b) to indicate the corresponding relative magnitudes of the two-component structure. }
\label{LPP scanning}
\end{figure*}

\begin{figure*}
\begin{minipage}[t]{0.32\textwidth}
\centering
\includegraphics[width=\textwidth]{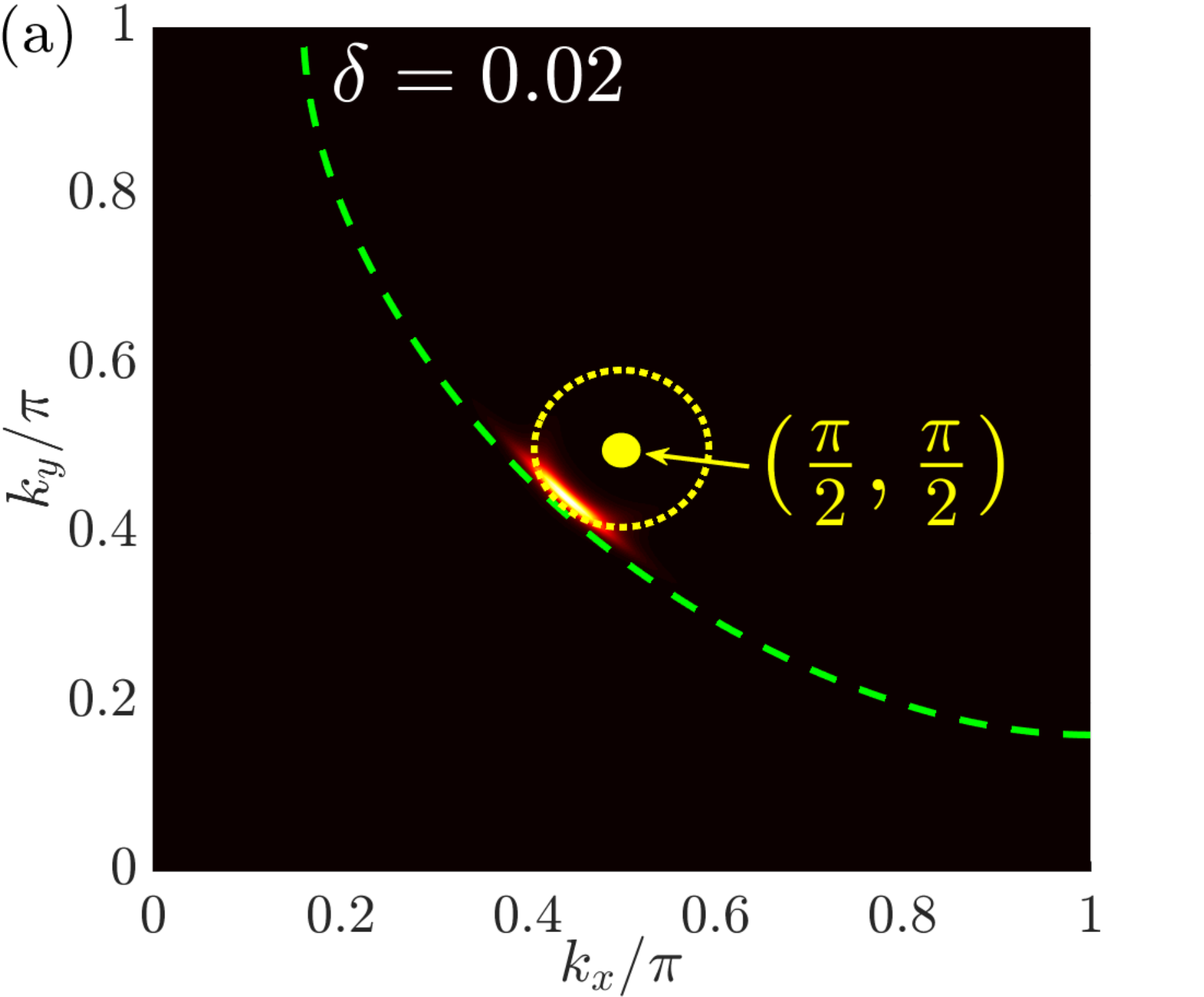}
\end{minipage}%
\begin{minipage}[t]{0.33\textwidth}
\centering
\includegraphics[width=\textwidth]{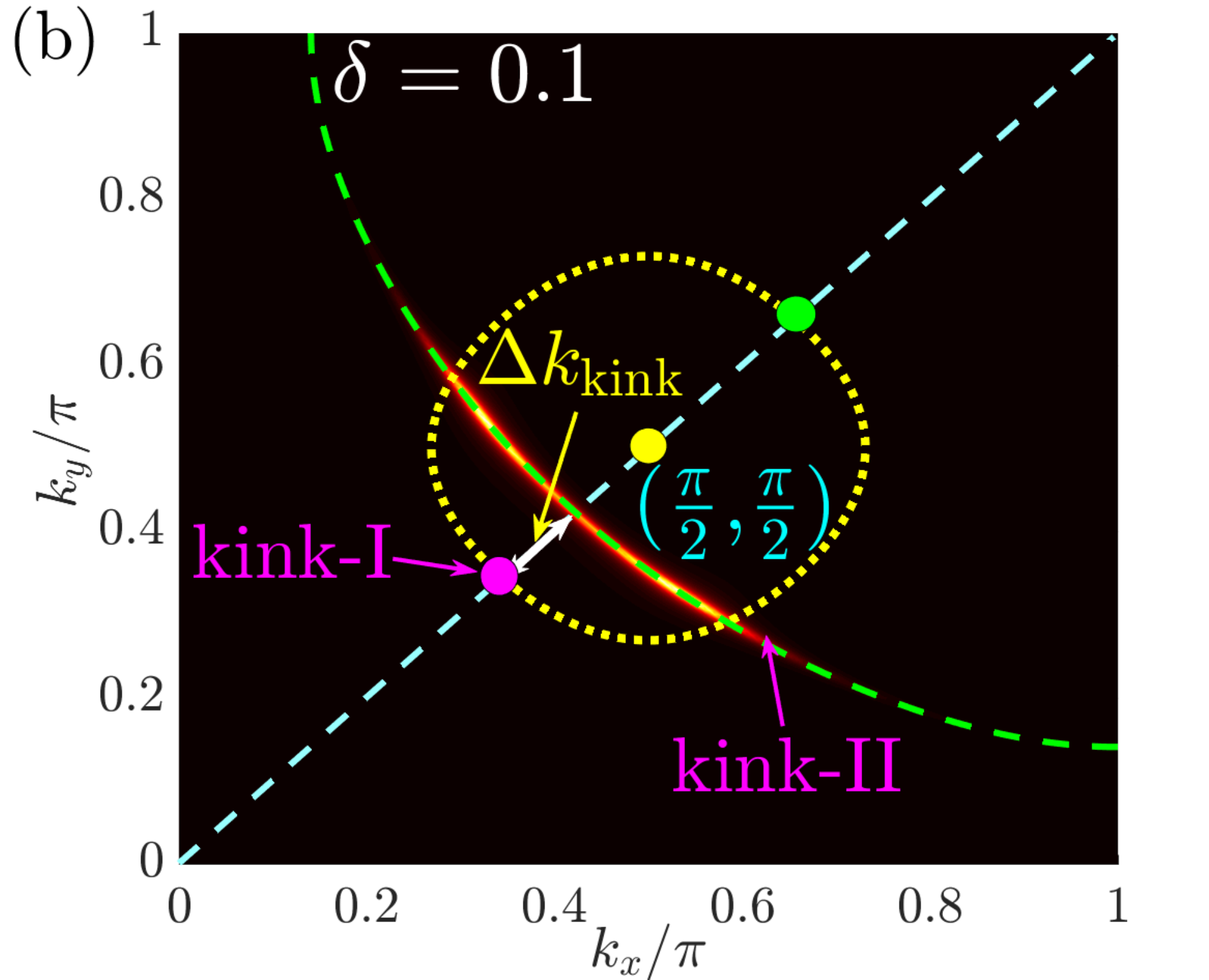}
\end{minipage}%
\begin{minipage}[t]{0.318\textwidth}
\centering
\includegraphics[width=\textwidth]{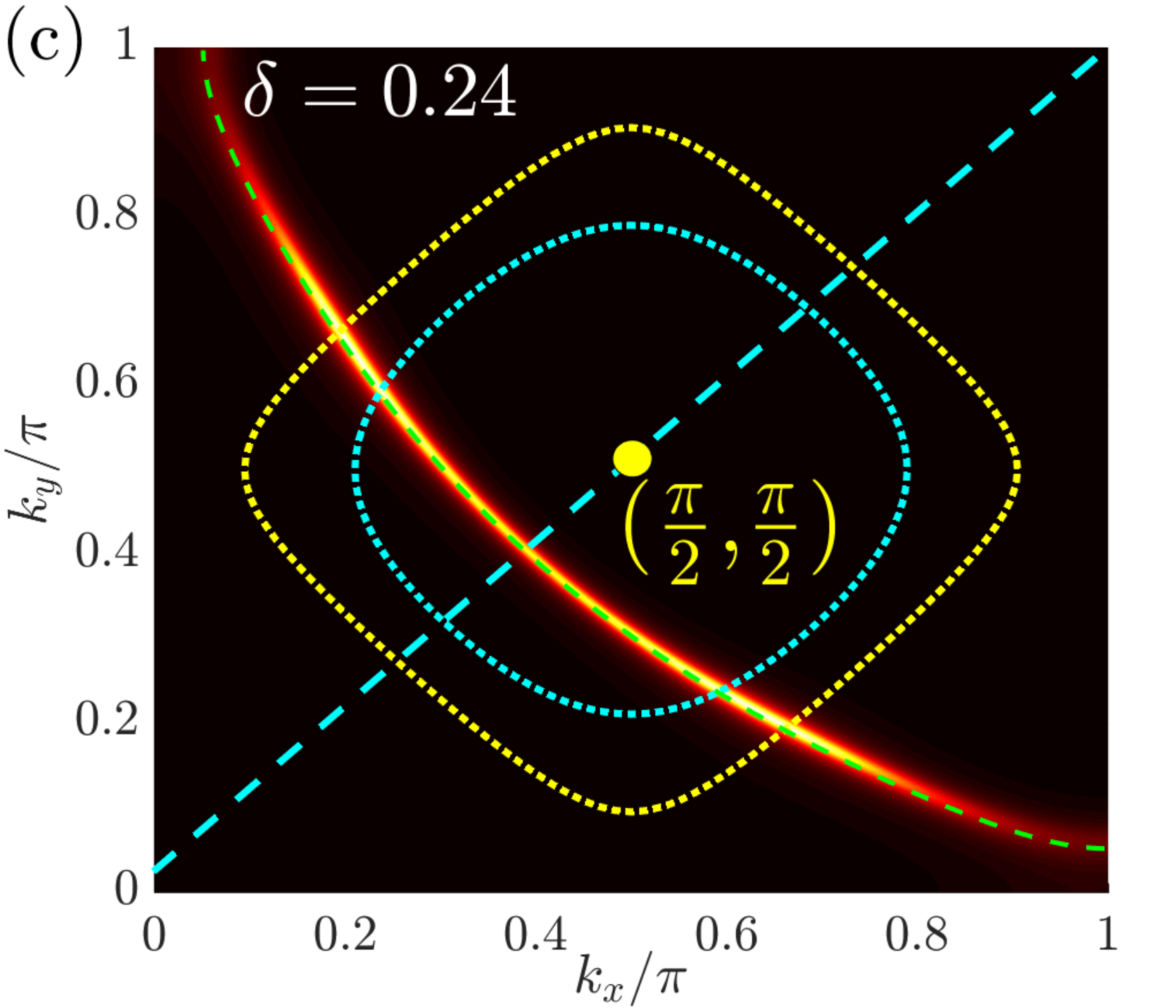}
\end{minipage}%
\caption{Fermi arc as represented by a sharp peak in $A(\bs{k},\omega=0)$ (bright yellow) in the LPP. Here the coherent quasiparticle emerges in $\hat{G}$, which coincides with the large Fermi surface of $\hat{G}_0$ but gets truncated roughly by the minimal energy contour [yellow dotted circle centred at $\bs{k}_0=(\pi/2,\pi/2)$] of the $a$-spinon at smaller doping (the inner dotted indigo circle at $\delta=0.24$ marks the corresponding Fermi pocket position, which becomes indistinguishable from the $s$-wave gap position at low doping). Such an ending point of the Fermi arc is marked by `kink-$\2$' in (b). Along the diagonal direction, there are another points intercepting the $a$-pocket, one is marked by `kink-$\1$' [violet bullet in (b)] towards the inside of the electron Fermi sea, while the other [green bullet in (b)] represents a high-energy ending point of the quasiparticle. The $a$-pocket provides a ``protection'' of the Fermi arc at low doping by the Pauli exclusion principle (see the text), whose effect is gradually diminished with the increase of $\Delta^a$ in the overdoping.}  
\label{protect}
\end{figure*}
\begin{figure*}[htbp]
\subfigure{
\begin{minipage}[p]{0.26\textwidth}
\centering
\includegraphics[width=\textwidth]{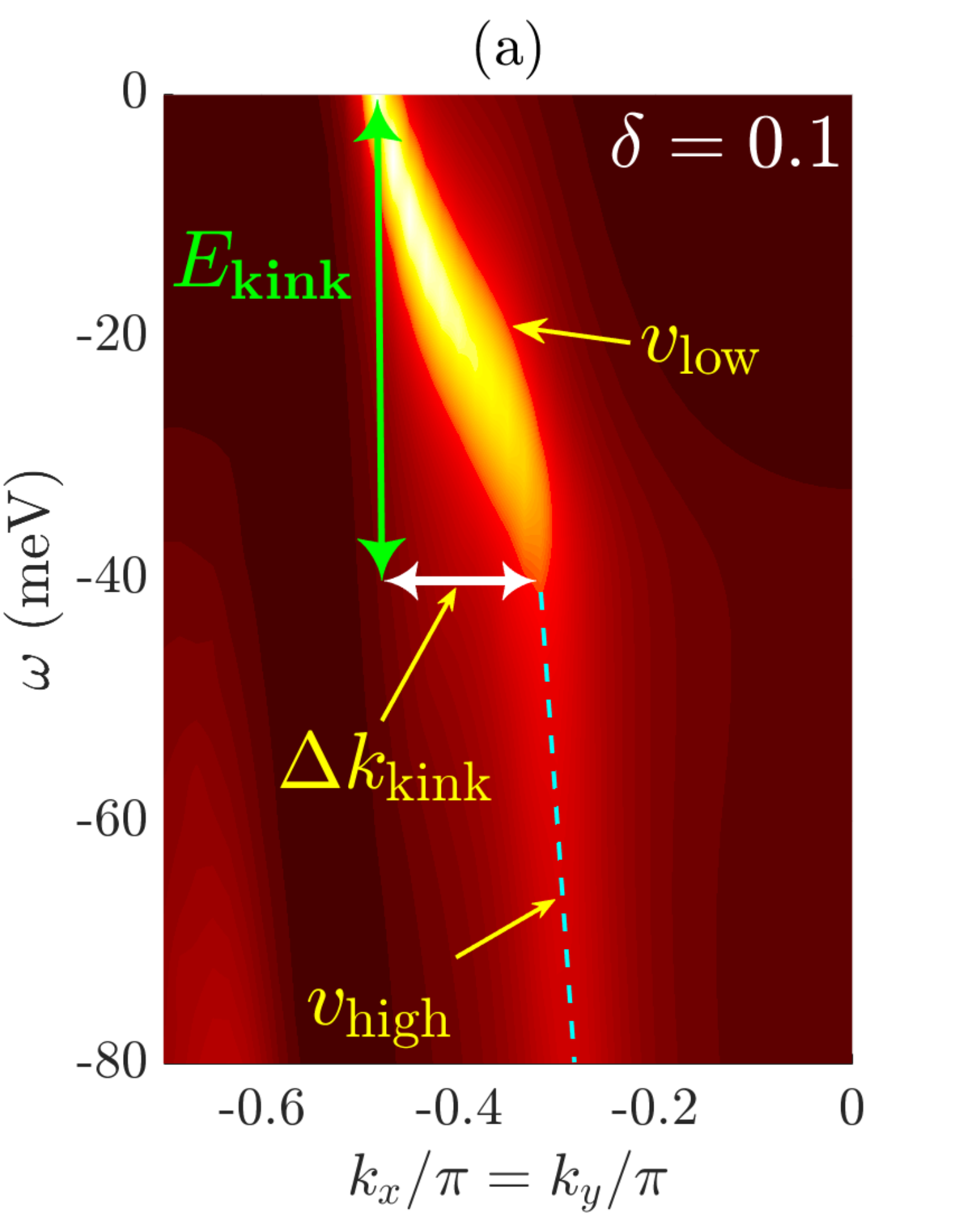}
\end{minipage}%
\begin{minipage}[p]{0.74\textwidth}
\centering
\includegraphics[width=\textwidth]{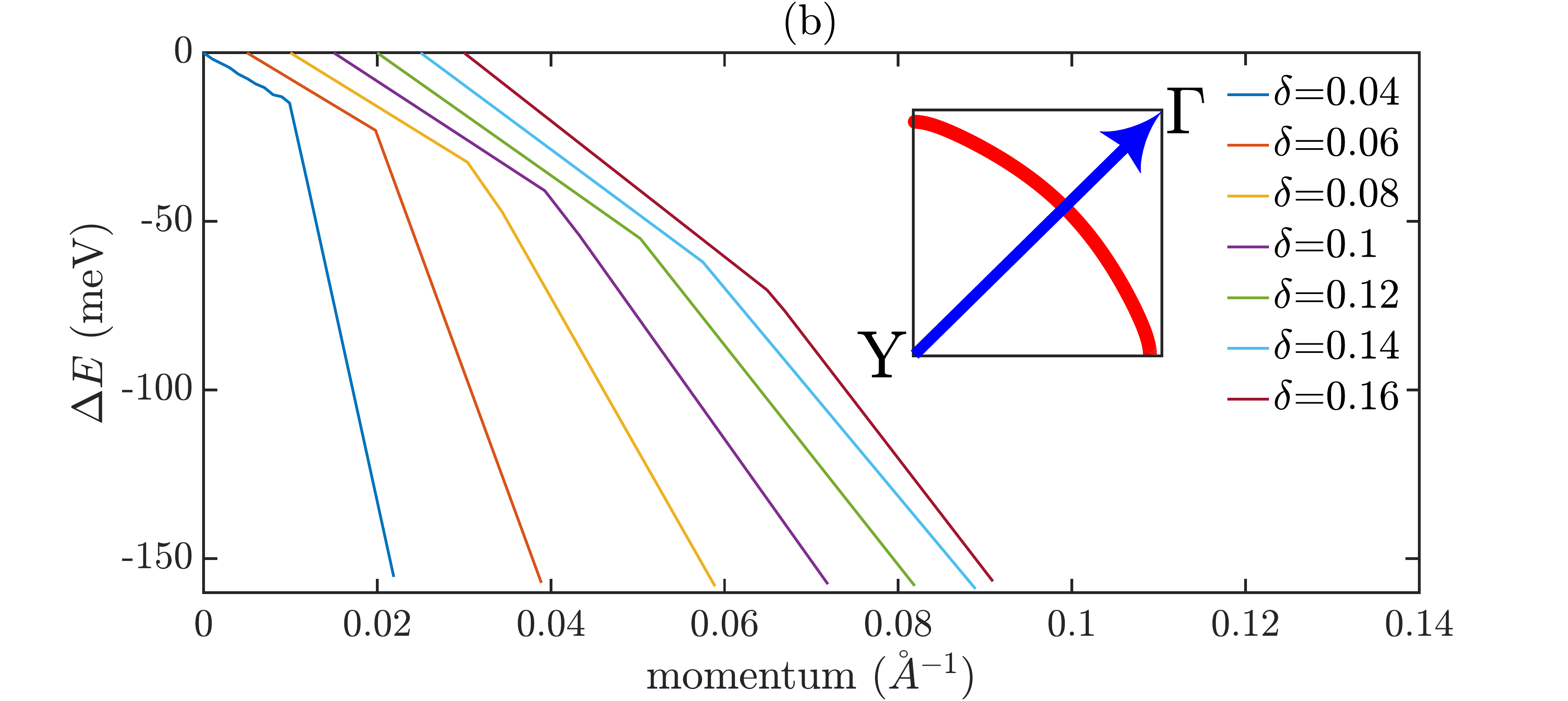}
\end{minipage}%
}
\subfigure{
\begin{minipage}[t]{0.6\textwidth}
\includegraphics[width=\textwidth]{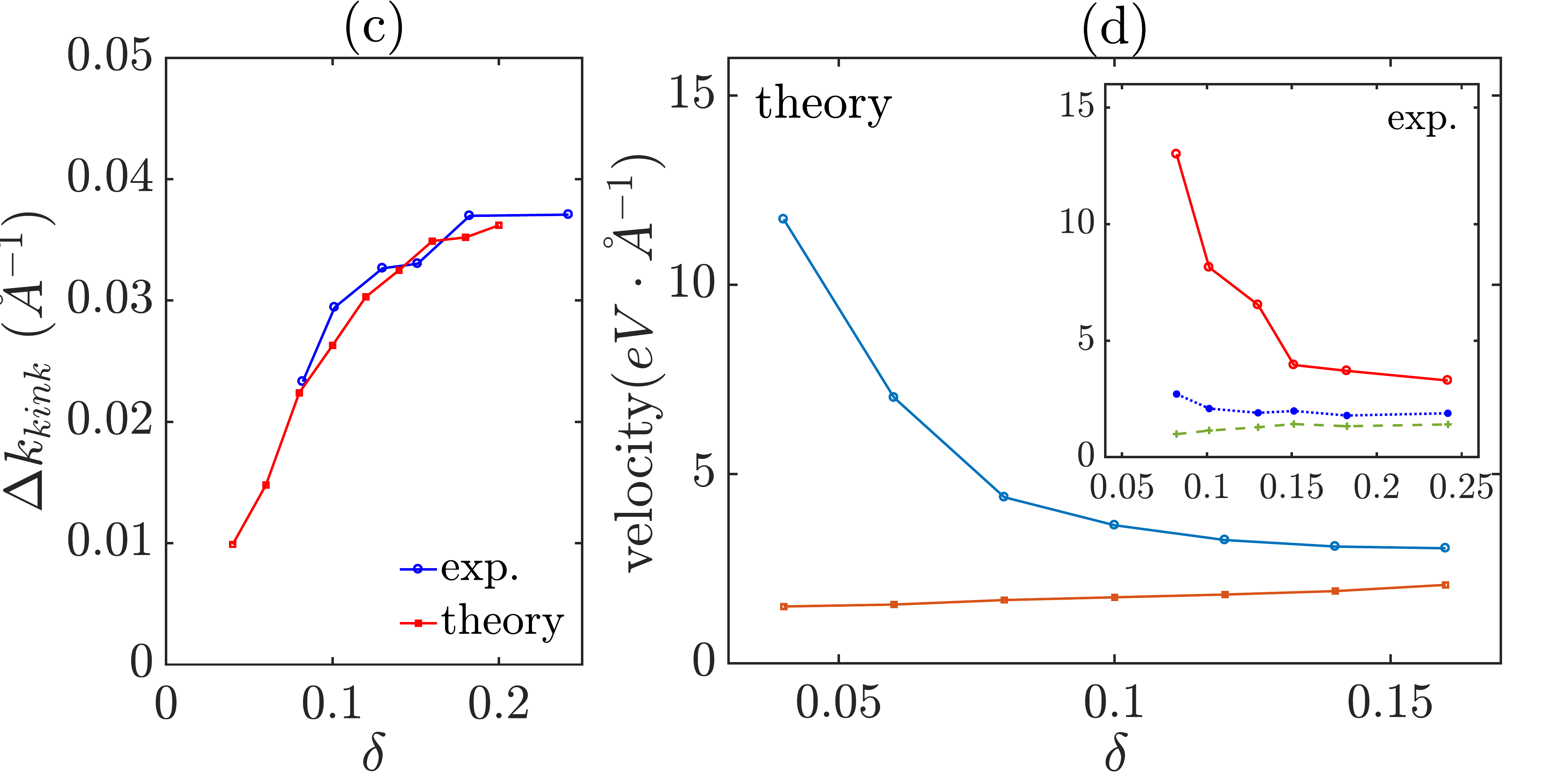}
\end{minipage}
\begin{minipage}[t]{0.39\textwidth}
\includegraphics[width=\textwidth]{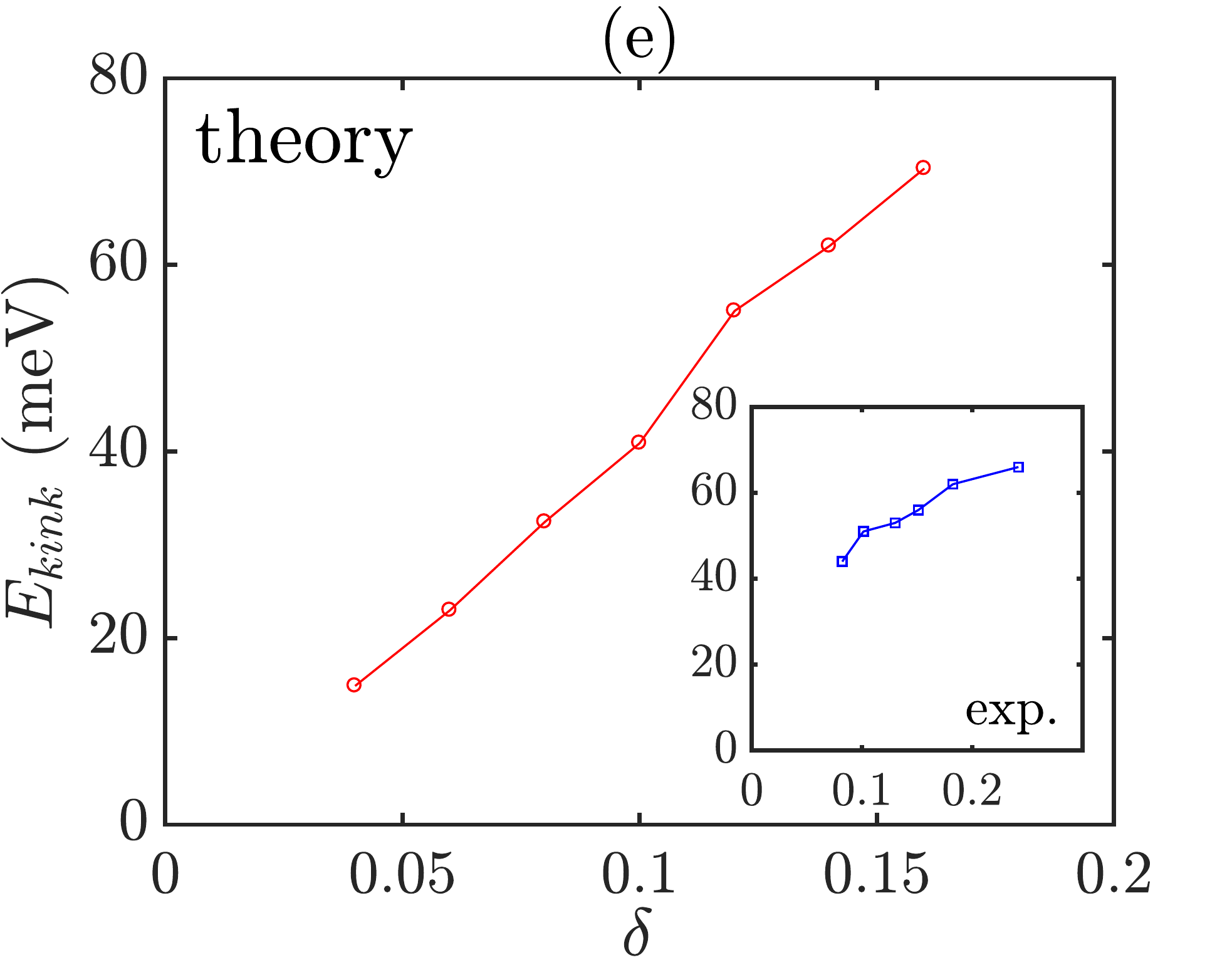}
\end{minipage}
}
\caption{(a): A typical ``kink'' in the dispersion of the quasiparticle peak in $A(\bs{k},\omega)$ along the nodal direction ($k_x=k_y$) [cf. the insert of (b)]; Its systematic evolution versus doping concentration is shown in (b); (c): $\Delta k_{\text{kink}}$ measuring the distance between the ``kink-$\1$'' and the Fermi point along the nodal direction [cf. Fig.\ref{protect}(b)]: theory (red) vs. experiment \cite{ZhongYG2019} (blue) as a function of doping $\delta$; (d): The Fermi velocities, $v_{\text{high}}$ and $v_{\text{low}}$, of high-energy (blue dot) and low-energy (red dot) modes [as marked in (a)] vs. doping. Insert: the experimental data \cite{ZhongYG2019}; (e): $E_{\text{kink}}$ vs. doping. Insert: the experiment \cite{ZhongYG2019}.} 
\label{nodalkink}
\end{figure*}

\begin{figure*}
\centering
\begin{minipage}[t]{0.33\textwidth}
\centering
\includegraphics[width=\textwidth]{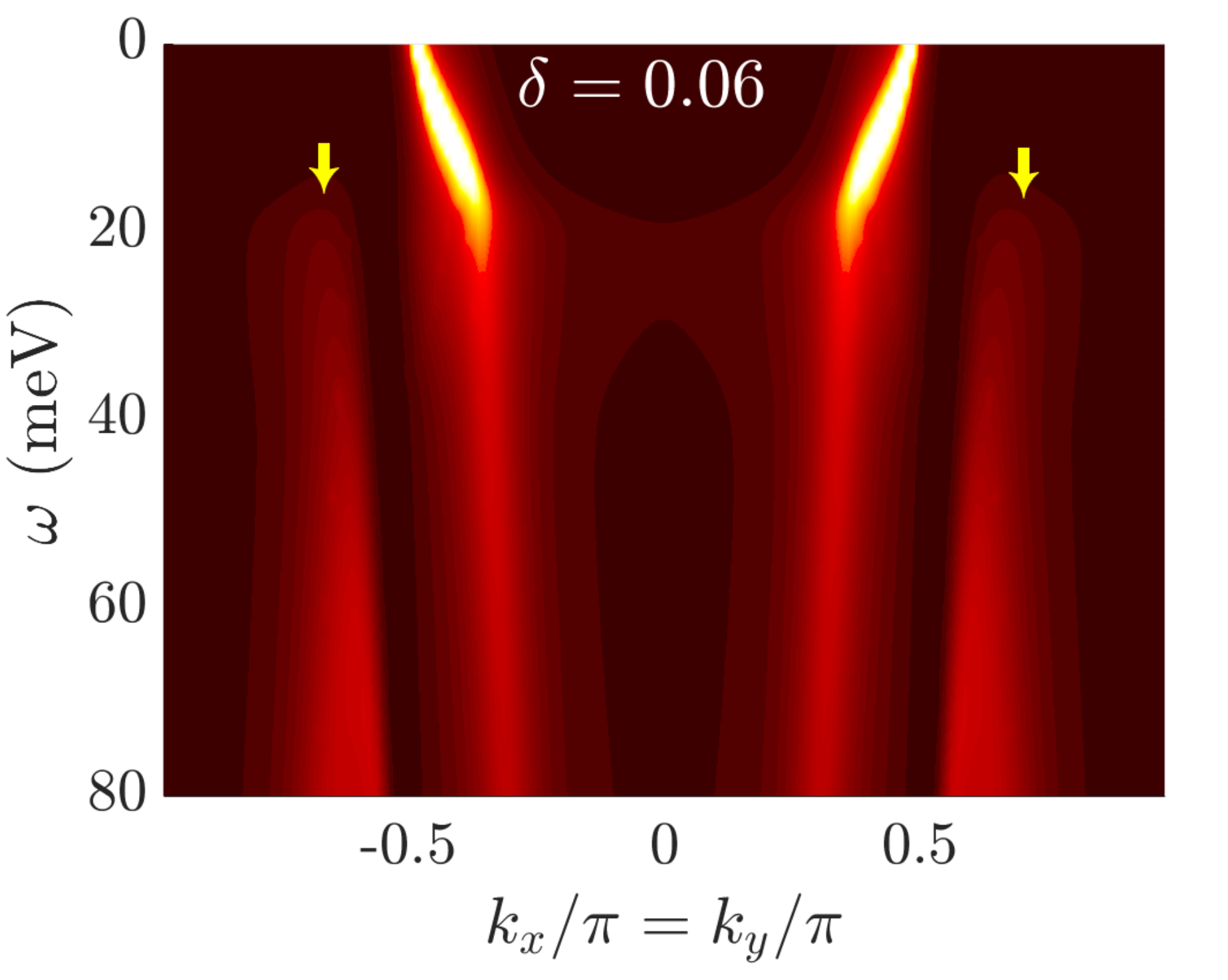}
\end{minipage}%
\begin{minipage}[t]{0.33\textwidth}
\centering
\includegraphics[width=\textwidth]{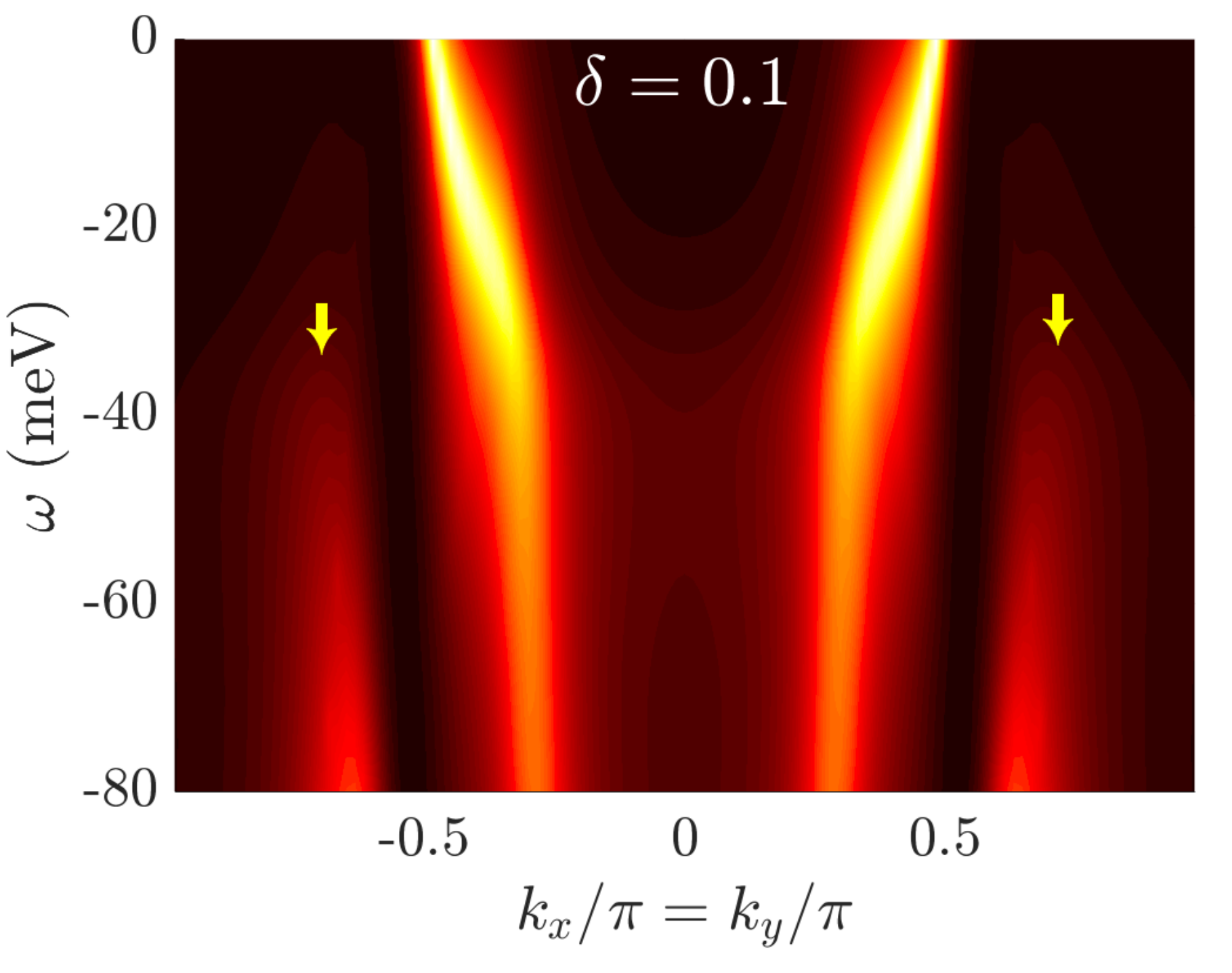}
\end{minipage}%
\begin{minipage}[t]{0.336\textwidth}
\centering
\includegraphics[width=\textwidth]{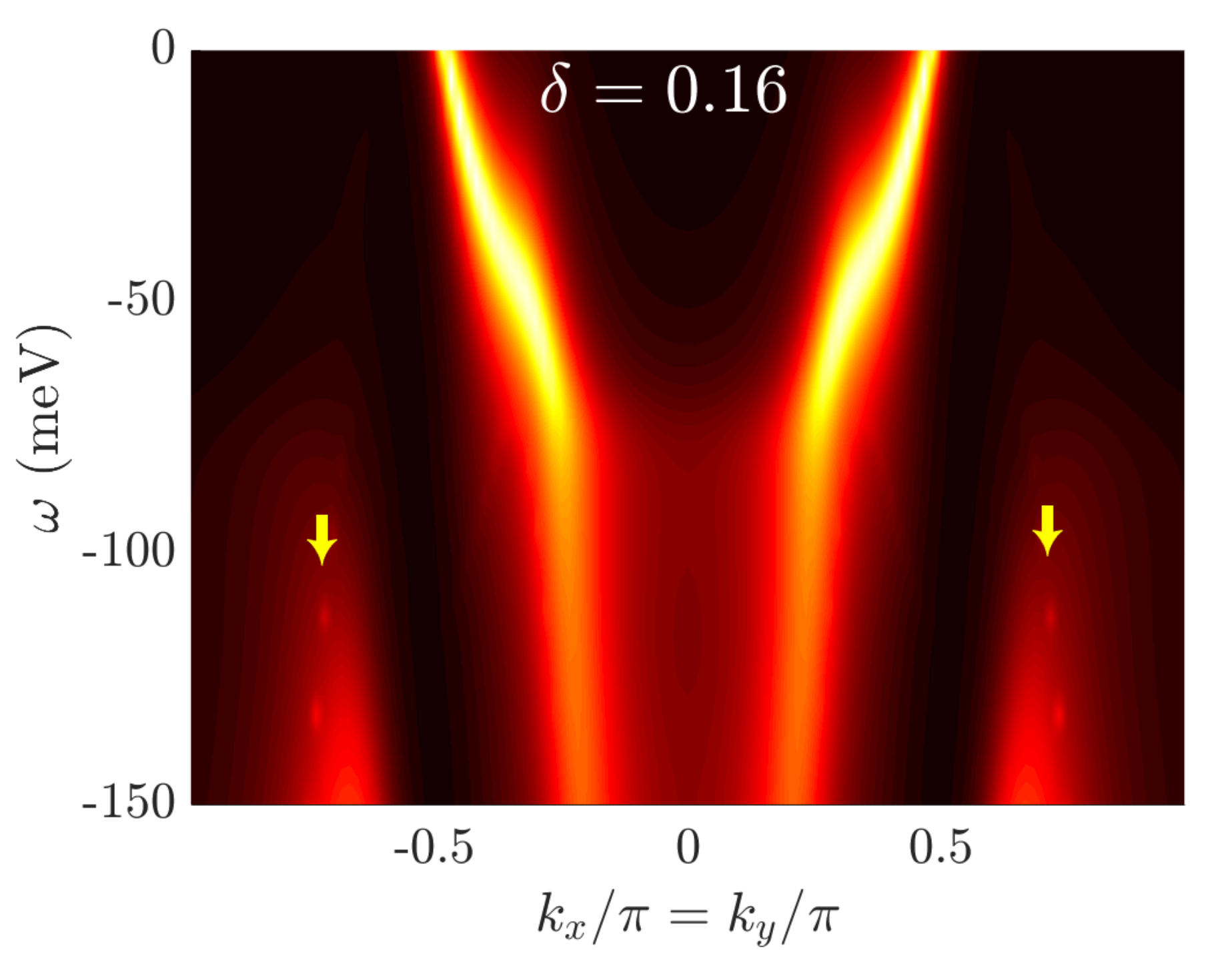}
\end{minipage}%
\caption{A systematic evolution of the spectral function along the nodal direction ($k_x=k_y$) with the whole range of momenta. Apart from the “kink-$\1$” phenomenon indicated in Fig. \ref{nodalkink}, an additional hidden high-energy mode is shown as marked by yellow arrows. As illustrated in Fig. \ref{protect}(b), such a mode corresponds to the fractionalization on the other side of the $a$-pocket along the nodal direction.}
\label{hidden}
\end{figure*}

\section{Spectral function: Experimental consequences\label{section4}}

One may determine the quasiparticle spectral function by:
\begin{align}
A(\bs{k},\omega)=-\frac{1}{\pi}\mathrm{Im}\Big(\big[\hat{G}(\bs{k},\omega+i0^+)\big]_{11}\Big)~.
\label{A}
\end{align}
Here the Green's function $\hat{G}$ has a two-component structure in Eq. (\ref{G3}), with the fractionalization component $\hat{\cal D}^0\sim \hat{D}^0 $ and the quasiparticle component $\hat{G}^0$. We have seen that $\hat{D}^0$ describes four Fermi pockets with an s-wave pairing as illustrated in Figs. \ref{Fermi pocket}(a) and (b). By contrast, $\hat{G}^0$ describes the Bogoliubov quasiparticle of the d-wave pairing $\Delta^{\text{SC}}$ with a large Fermi surface determined by the band structure satisfying the Luttinger volume of the total electrons. In the following, we first examine the basic features of $A(\bs{k},\omega)$ in the LPP.

\subsection{Emergent Fermi arc due to fractionalization in the LPP}
In the LPP, the SC order parameter $\Delta^{\text{SC}}=0$ such that a \textit{full} large Fermi surface is expected to recover in $\text{Im}(\hat{G}^0)$ at $\omega=0$. $A(\bs{k},\omega)$ at $\omega=0$ can be directly calculated at different doping concentrations as presented in Fig. \ref{arc}, where a ``Fermi arc'' structure naturally appears in the underdoped LPP.

Here, the segments of the Fermi surface, i.e., the Fermi arcs, in Fig. \ref{arc} at $\omega\simeq 0$ are protected by the $s$-wave gapped Fermi pockets of the fractionalized $\hat{\cal D}^0$ whose spectral function is gapped. In Fig. \ref{LPP scanning}(a), a two-component structure in $A(\bs{k},\omega)$ has been shown by projecting onto the k-space, where the Fermi pockets from $\hat{\cal D}^0$ at $\omega\sim 50 $ $meV$ are much broadened as composed to Fig.~\ref{Fermi pocket} for $\hat{D}^0$ by introducing a phenomenological $f_0$ discussed in Appendix \ref{App.B}. (But the broadening essentially has no affect on the Fermi arcs at $\omega=0$, which is within the gap.) Here the momentum is scanned along the bare large Fermi surface in the first quarter of the Brillouin zone at $\delta=0.1$ (cf. the inset of Fig. \ref{LPP scanning}(b)). Energetically there is always an $s$-wave pairing gap opening up in the Fermi pocket of $\hat{D}^0$ in the upper branch of the spectral function, while a gapless lower branch emerges along the large Fermi surface of $\hat{G}^0$ in the LPP [cf. Fig. \ref{LPP scanning}(c) for the relative strengths in a 3D plot]. 

In Fig. \ref{protect}, the Fermi pocket and the large bare Fermi surface of $\hat{D}^0$ are presented at three typical dopings in the one quarter of the first Brillouin zone. It illustrates that the large Fermi surface gets truncated by the Fermi pocket to result in the “Fermi arc” at small doping [cf. Figs. \ref{protect} (a) and (b)] with the arc termination point marked by “kink-$\2$” in Fig. \ref{protect}(b). On the other hand, in an overdoped case where the BCS-like pairing $\Delta^a$ is sufficiently large (cf. Appendix \ref{App.A}), the Fermi arc can well extend outside the Fermi pocket as indicated in Fig. \ref{protect} (c) at $\delta=0.24$. Note that the minimal energy at a finite $\Delta^a$ (yellow dotted circle) becomes increasingly larger than the Fermi pocket at $\Delta^a=0$ (indigo dotted circle) as shown in Fig. \ref{protect} (c), while they approximately coincide at smaller doping [not shown in Figs. \ref{protect}(a) and (b)].

Physically, to leading order of approximation, a quasiparticle should generally decay into an $a$-spinon according to Eq. (\ref{fractionalization}) as described by $\hat{\cal D}^0$. However, the Fermi pockets of the $a$-spinon in the ground state (\ref{ground state wavefunction}) will protect the quasiparticle from decaying inside the pockets due to the Pauli exclusion principle. In other words, the missing portions of the large Fermi surface in the ARPES experiment observed in the pseudogap phase of the cuprate can be naturally explained by the electron fractionalization outside the Fermi arc segments, while the Fermi arc itself may be regarded as the emergent quasiparticle protected at $\omega=0$, which is roughly within the $a$-spinon pockets at low doping as shown in Fig. \ref{protect}. With the increase of doping, the enlarged $\Delta^a$ means that the gapped $a$-spinon pocket is pushed further away from $\omega=0$ to weaken the fractionalization effect of the quasiparticle to decay into the $a$-spinon.

\subsection{Nodal-direction kink due to fractionalization}
As shown by Fig. \ref{protect}(b), the quasiparticle excitation can still maintain its coherence within the Fermi pocket (or an $a$-pocket as it is formed by the $a$-spinons). This is not only true along the large Fermi surface (Fermi arc), but also valid along a nodal direction that connects, say, $\Gamma=(0,0)$ and $Y = (\pi,\pi)$ in Fig. \ref{protect}. Note that along this direction, the spectral function remains essentially the same for both LPP and SC phase as the $d$-wave SC gap vanishes.

The spectral function indicates a dispersion of the quasiparticle excitation with a velocity $v_{\text{low}}$ along the nodal direction inside the $a$-pocket, and then a ``kink'' with much reduced spectral weight and larger velocity $v_{\text{high}}$ outside the $a$-pocket [marked by “kink-$\1$” in Fig. \ref{protect}(b)], as shown in Fig. \ref{nodalkink}(a). A systematic evolution of the kink energy $E_{\text{kink}}$ with doping concentration is shown in Fig. \ref{nodalkink}(b).

Some more detailed features with the experimental comparisons are given in Figs. \ref{nodalkink}(c)-(e):
\begin{enumerate}[1.]
\item According to the definition in Fig. \ref{protect}(b), $\Delta k_{\text{kink}}$ measures the distance between the Fermi arc and the kink-$\1$ along the nodal direction. It increases monotonically with doping [red curve in Fig. \ref{nodalkink}(c)] in an excellent agreement with the ARPES data \cite{ZhongYG2019} (blue curve);
\item The Fermi velocities, $v_{\text{low}}$ and $v_{\text{high}}$, of the low-energy and high-energy modes as functions of doping are shown in the main panel of Fig. \ref{nodalkink}(d), while the experimental ARPES data \cite{ZhongYG2019} are presented in the inset for comparison. Similar behavior has been also previously observed in the ARPES experiment of Ref. \onlinecite{Zhou2003};
\item A systematic doping dependence of $E_{\text{kink}}$ is shown in the main panel of Fig. \ref{nodalkink}(e), while the experimental data \cite{ZhongYG2019} are plotted in the insert.
\end{enumerate}

\begin{figure*}
\subfigure{
\centering
\begin{minipage}[t]{0.48\textwidth}
\centering
\includegraphics[width=\textwidth]{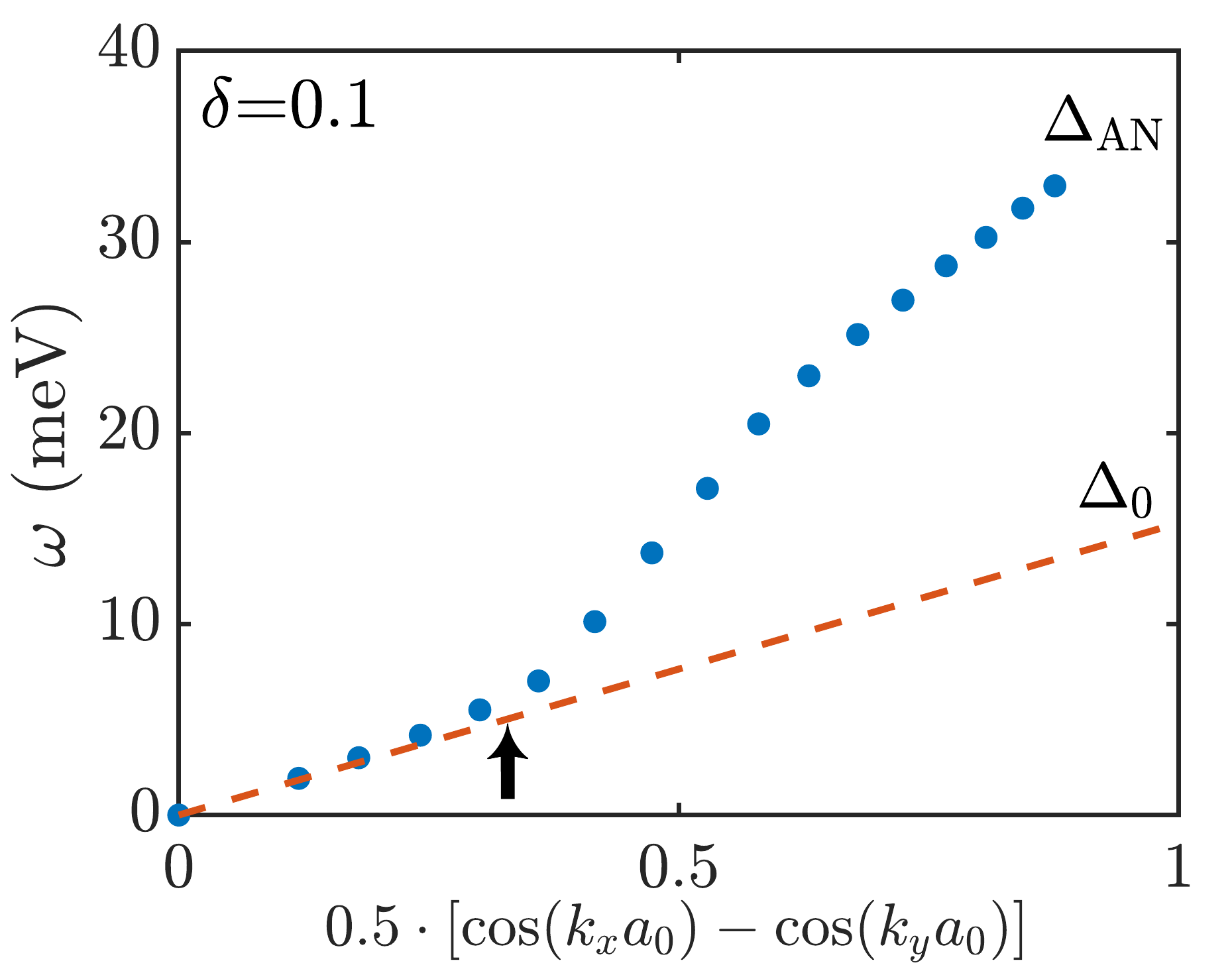}
\end{minipage}%
\begin{minipage}[t]{0.48\textwidth}
\centering
\includegraphics[width=\textwidth]{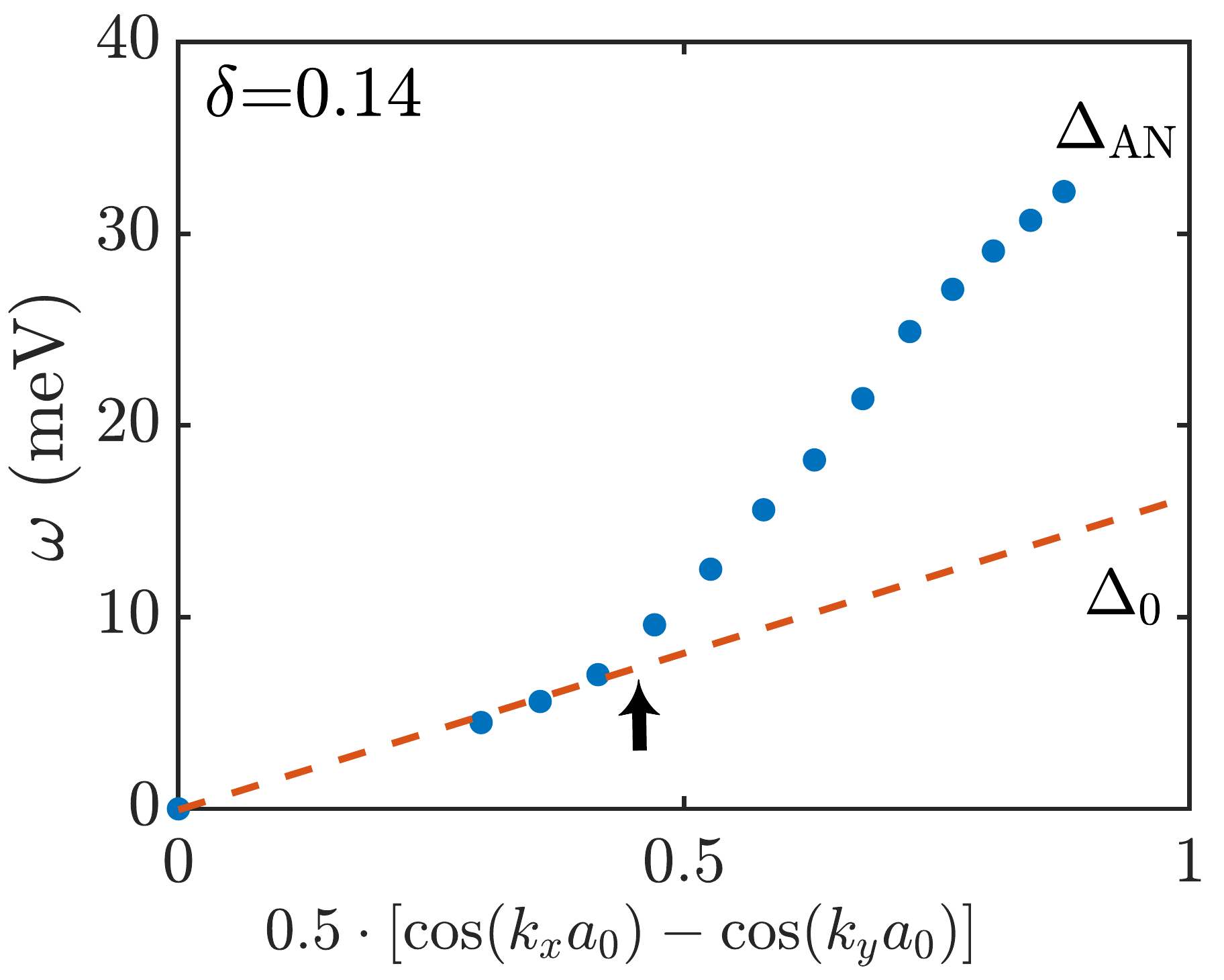}
\end{minipage}%
}
\subfigure{
\begin{minipage}[t]{0.48\textwidth}
\centering
\includegraphics[width=\textwidth]{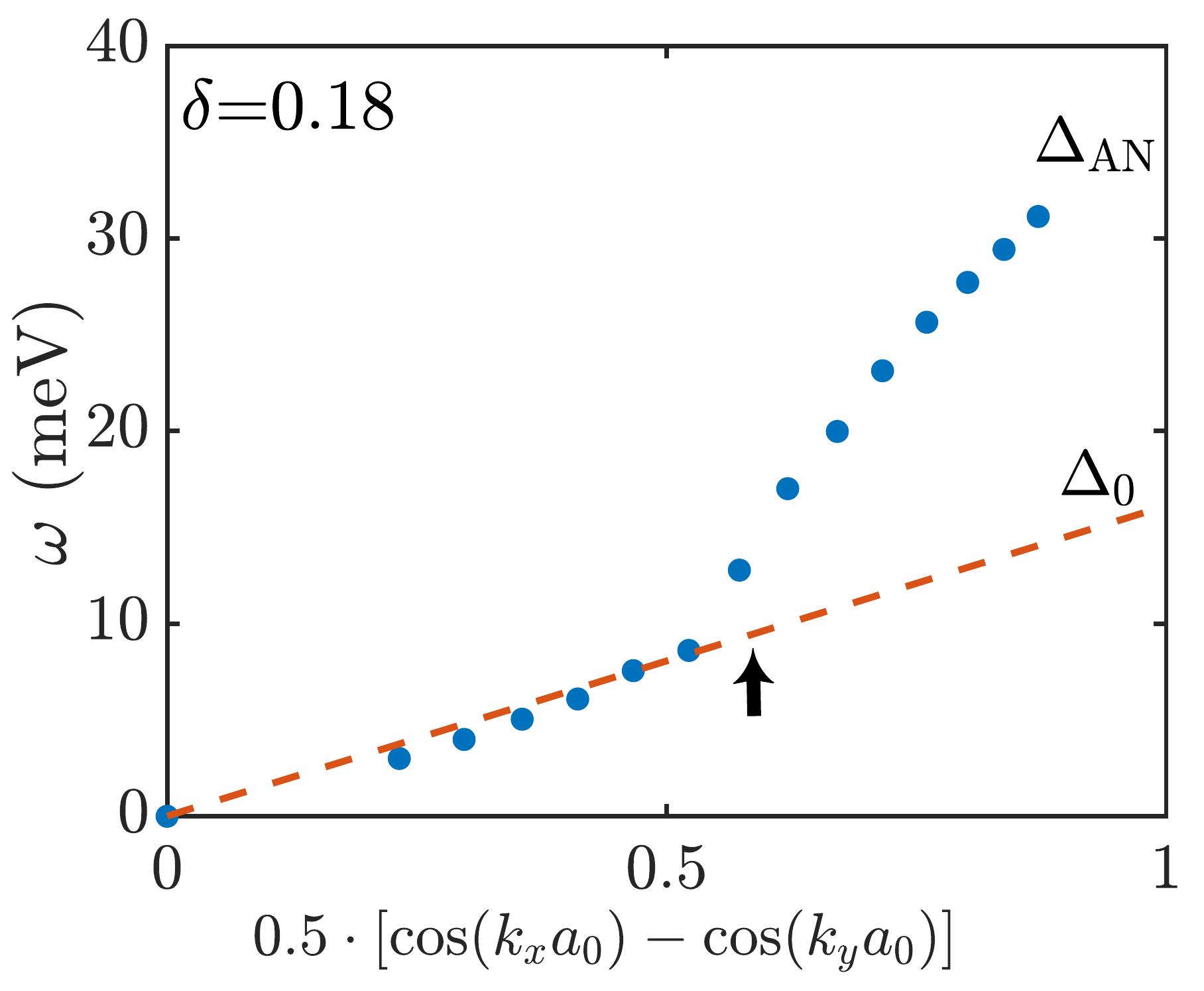}
\end{minipage}%
\begin{minipage}[t]{0.48\textwidth}
\centering
\includegraphics[width=\textwidth]{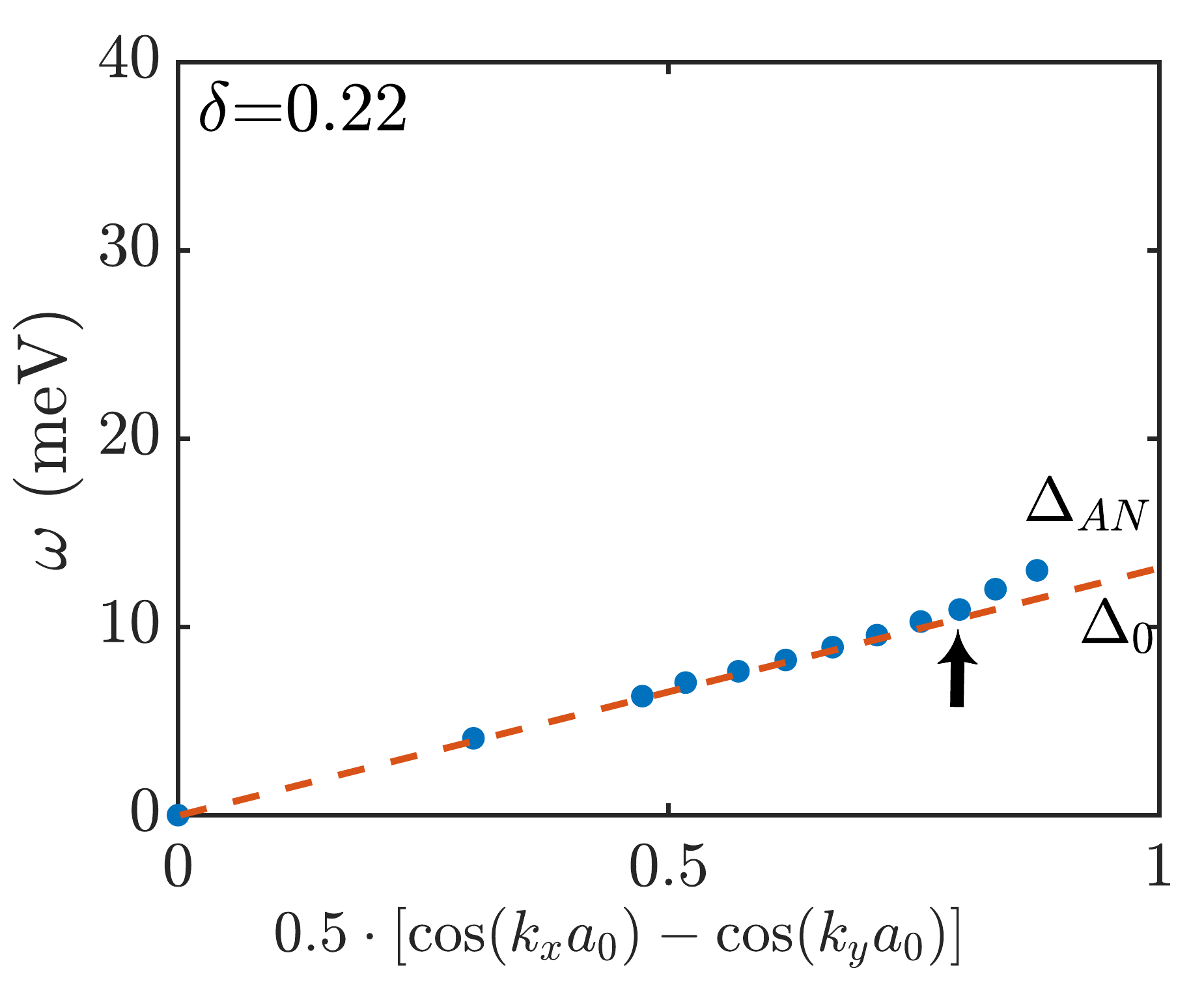}
\end{minipage}%
}
\caption{Two-gap structure in the SC phase: a small $d$-wave gap is opened up along the Fermi arc (cf. the low-energy branch in Fig. \ref{LPP scanning}), which is determined by a low-lying sharp peak in $A(\bs{k},\omega)$ with $\bs{k}$ scanning perpendicular to the Fermi surface (cf. the insert of Fig. \ref{LPP scanning} at a given $\theta$) and is plotted against $0.5[\cos(k_xa_0)-\cos(k_ya_0)]$ with an extrapolation to $\Delta_0$ at the Brillouin zone boundary. An arrow marked the ending point of the Fermi arc, beyond which a new sharp mode emerges in $A(\bs{k},\omega)$ with a larger gap $\Delta_{\text{AN}}$ in the antinodal regime.} 
\label{angle}
\end{figure*}

One finds a very consistent and overall quantitative agreement between the theory and experiment. We emphasize that no fitting has been made since the calculation of the spectral function is based on the previous mean-field solution \cite{MaYao2014}. One may interpret the excellent agreement as due to the fact that the main features are all determined by the Fermi packet size of the $a$-spinon, which is solely determined by doping concentration $\delta$.

In the limit of $\delta\rightarrow0$, the Fermi pockets will shrink into four momenta at $\bs{k}_0=(\pm\pi/2,\pm\pi/2)$. A variational Monte Carlo calculation of the ground state wavefunction based on the same fractionalization given in Eqs. (\ref{fractionalization}) and ({\ref{ground state wavefunction}}) has shown that the quasiparticle excitation only exists at $\bs{k}_0$ with a vanishing quasiparticle spectral weight $Z_{\bs{k}_0}$ by a finite-size scaling, while the rest of the momentum distribution is indeed contributed by the fractionalized particle $\tilde{c}_{i\sigma}=\hat{c}_{i\sigma}e^{-i\hat\Omega_i}=h_i^\dag a_{i\bar\sigma}^\dag$ in the single-hole ground state, which are in excellent agreement with the numerical DMRG results.

\textit{Prediction: a hidden high-energy mode.} Apart from the high-energy fractionalized mode contributing to the ``kink'' phenomenon around the kink-$\1$ along the nodal direction, another ``kink'' point is also predicted at the $a$-pocket closer to the $Y$ point along the $\Gamma-Y$ line, which is marked by the green dot in Fig. \ref{protect}(b). Such an additional high-energy mode is shown in Fig. \ref{hidden} (indicated by an short yellow arrow) for various doping concentrations, which is increasingly strong with reducing doping, and remains to be observed experimentally by ARPES as a unique prediction of the present fractionalization theory.

\begin{figure*}
\begin{minipage}[t]{0.246\textwidth}
\centering
\includegraphics[width=\textwidth]{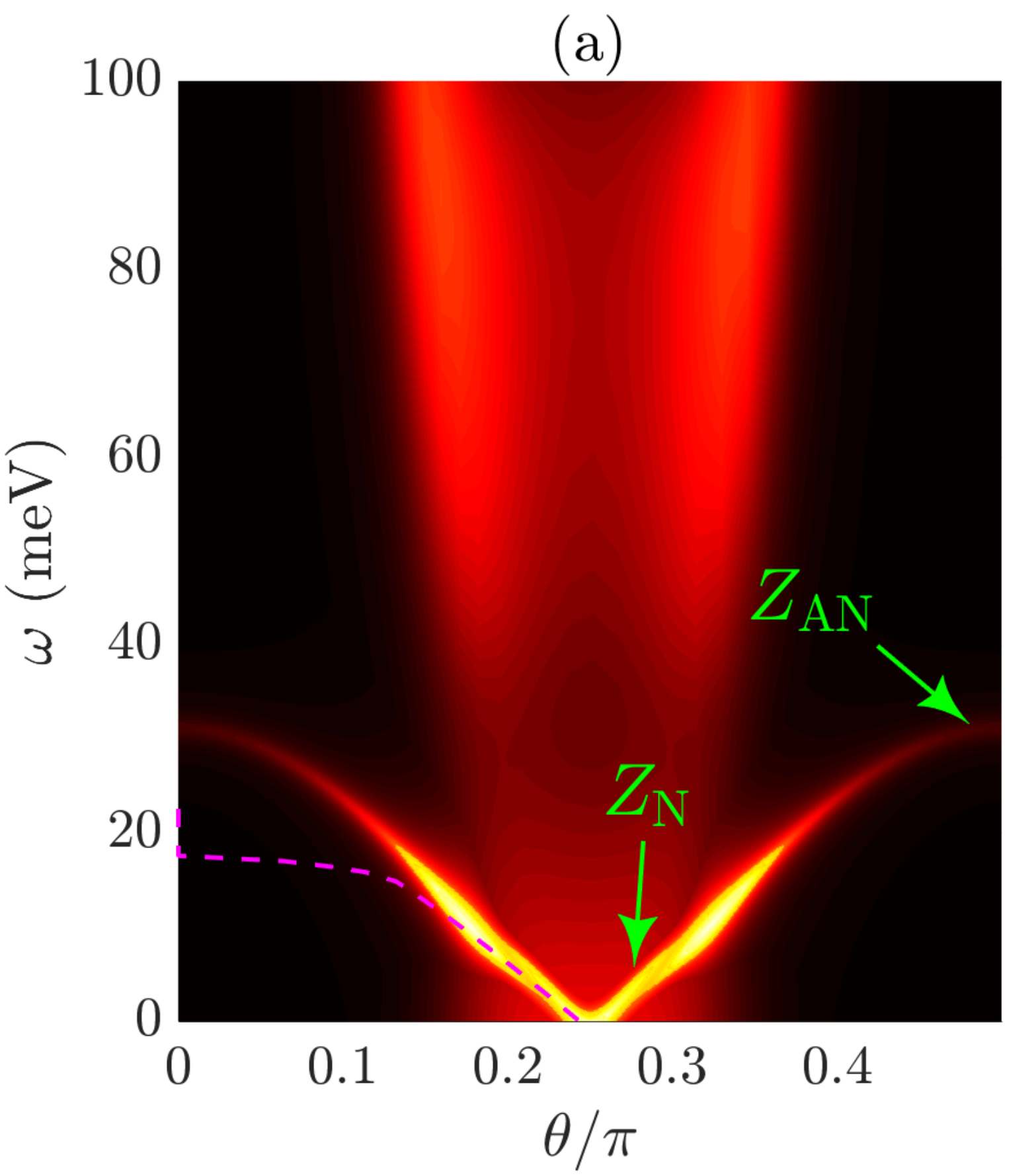}
\end{minipage}%
\begin{minipage}[t]{0.36\textwidth}
\centering
\includegraphics[width=\textwidth]{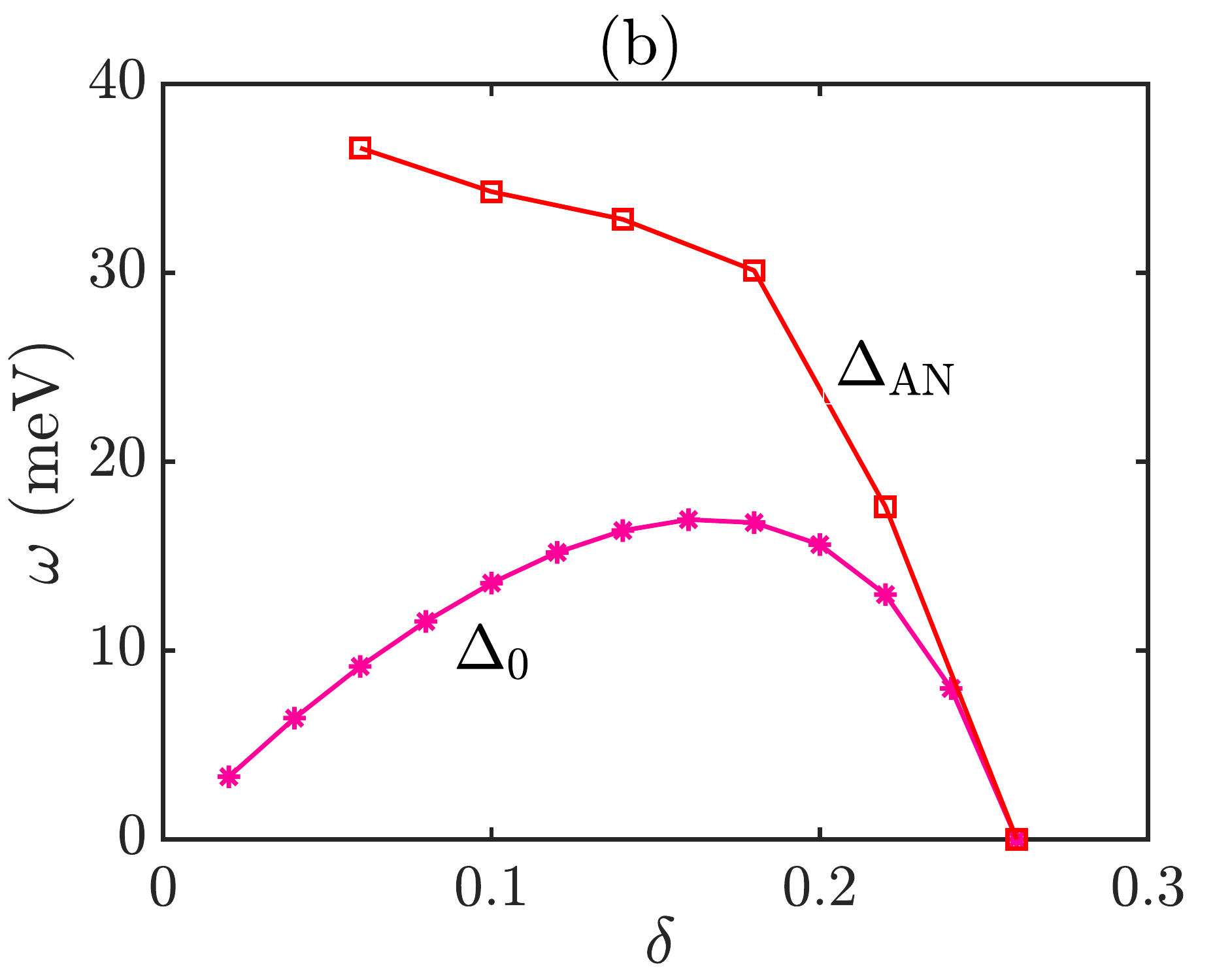}
\end{minipage}%
\begin{minipage}[t]{0.382\textwidth}
\centering
\includegraphics[width=\textwidth]{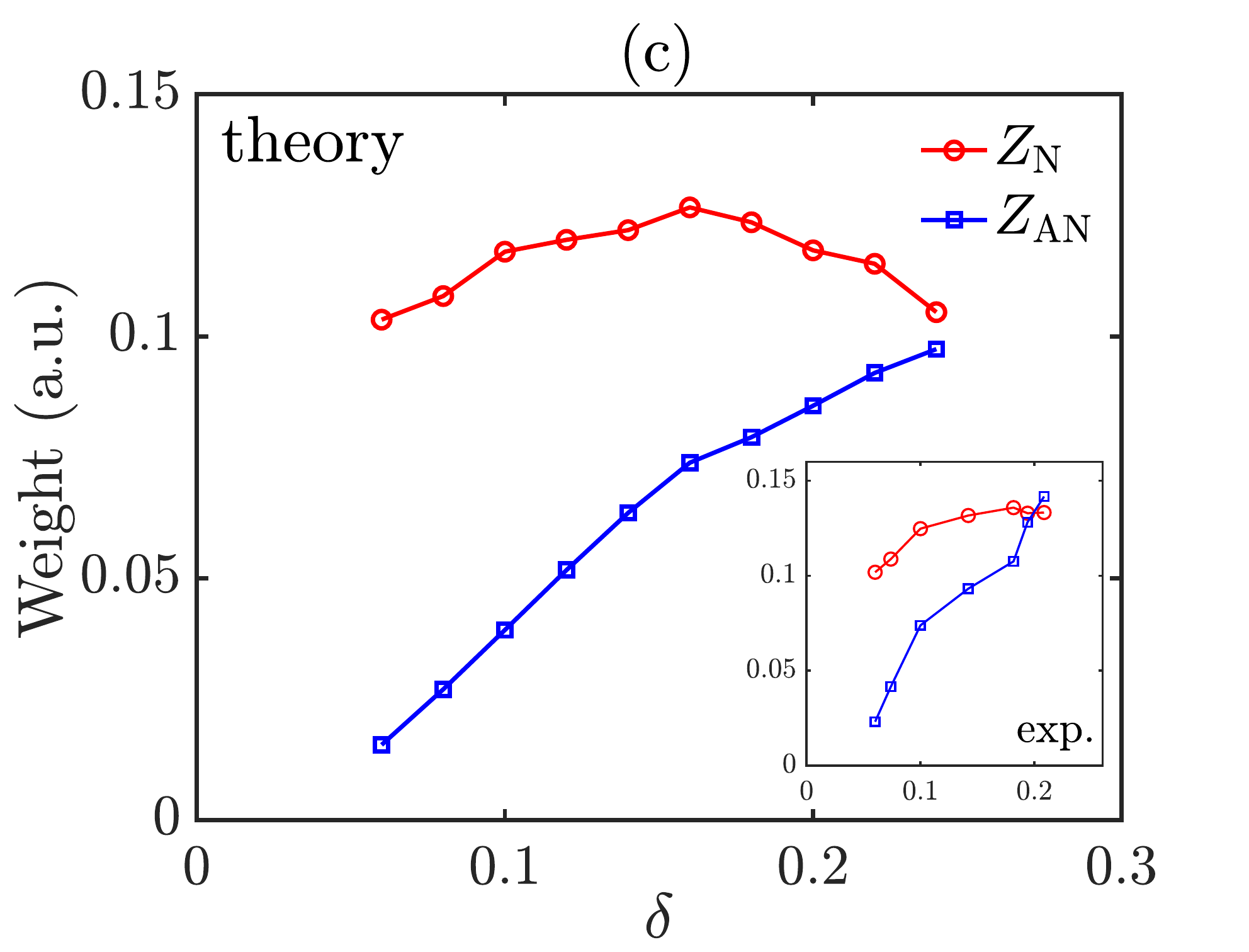}
\end{minipage}%
\caption{(a): The two-gap structure of the spectral function in the SC phase with the ${\bf k}$-scanning along the bare Fermi surface, in which the nodal and antinodal regimes are characterized by the quasiparticle spectral weights $Z_{\text{N}}$ and $Z_{\text{AN}}$, respectively. The dashed line indicates a bare $d$-wave gap function; (b): The corresponding two-gaps, $\Delta_0$ and $\Delta_{\text{AN}}$, versus the doping concentration, which eventually emerge and gradually vanish in the overdoped regime; (c): $Z_{\text{N}}$ varies placidly with respect to the doping concentration, while the quasiparticle weight $Z_{\text{AN}}$ of antinodal excitation increases monotonically as a function of $\delta$. Both calculated quantities (red dots) are in qualitative agreement with the experiment shown in the insert.
}
\label{position}
\end{figure*}

\subsection{Superconducting phase: A two-gap and new ``kink'' structure\label{4C}}
The SC phase is characterized by a simple $d$-wave order parameter $\Delta^{\text{SC}}\ne0$ given in Eq. (\ref{SCgap}), which will enter the full Green function $\hat{G}$ in Eq. (\ref{G}) through the bare hole Green’s function $\hat{G}_0$ in Eq. (\ref{G0}).

One can read off the low-energy sharp peak in the spectral function $A(\bs{k},\omega)$ with $\bs{k}$ scanning across the bare large Fermi surface at each scanning angle shown in the insert of Fig. \ref{LPP scanning}. In contrast to a novel Fermi arc feature at $\omega=0$ in the LPP, a gap feature will thus emerge in the SC phase associated with the Bogoliubov quasiparticle. By substituting Eq. (\ref{SCgap}) obtained by the mean-field theory, the resulting gap structure is presented in Fig. \ref{angle} at different doping concentrations. Instead of a single $d$-wave gap appearing in $\hat{G}_0$, such a gap structure as measured by $A(\bs{k},\omega)$ along the Fermi surface is found to generally break into two-gap structure with a new kink feature marked by an arrow in each panel of Fig. \ref{angle}.

In the following, we examine in detail how such a Bogoliubov quasiparticle emerges inside/outside the gap and pocket of the $a$-spinon spectrum in Figs. \ref{LPP scanning} and \ref{protect}(b). As indicated by the arrow in Fig. \ref{angle}, the ``kink'' (the ending point of the low-lying nodal excitation) coincides with the ending position of the Fermi arc in the LPP, which at smaller doping is marked by “kink-$\2$” in Fig. \ref{protect}. Hence, the $a$-pockets in the momentum space still plays an essential role for the observable kink effect on the Bogoliubov quasiparticle spectrum.

The two-gap structure in Fig. \ref{angle} is characterized by $\Delta_0$ and $\Delta_{\text{AN}}$ in front of the $d$-wave factor $0.5|\cos(k_xa_0)-\cos(k_ya_0)|$, which are determined by extrapolating to the Brillouin zone boundary. Such a kink/two-gap structure in the Bogoliubov quasiparticle spectrum separates the large Fermi surface into two parts, the nodal and antinodal regimes, as schematically summarized in Fig. \ref{position}(a). The systematic doping dependences of $\Delta_{\text{AN}}$ and $\Delta_0$ (which are labeled in each panel of the Fig. \ref{angle}) are shown in Fig. \ref{position}(b), which are in overall agreement with the experiment. Note that the distinction between $\Delta_0$ and $\Delta_{\text{AN}}$ decreases monotonically with the increase of doping concentration, which eventually disappears together with the ``kink'' feature in the overdoped regime [cf. also Fig. \ref{angle}].

The corresponding peak weights in $A(\bs{k},\omega)$ are denoted by $Z_{\text{N}}$ and $Z_{\text{AN}}$, respectively, at two representative points marked in Fig. \ref{position}(a) in the two-gap regimes. In the panel (c) of Fig. \ref{position}, the extracted values of the quasiparticle weight of the nodal excitation $Z_{\text{N}}$ and the quasiparticle weight of the anti-nodal excitation $Z_{\text{AN}}$ are presented, in comparison with the experimental data. $Z_{\text{N}}$ varies placidly with respect to the increasing of doping concentration, while $Z_{\text{AN}}$ increases monotonically and more drastically. Both are in quantitative agreement with the experiment.

\begin{figure*}
\begin{minipage}[h]{0.33\textwidth}
\includegraphics[width=\textwidth]{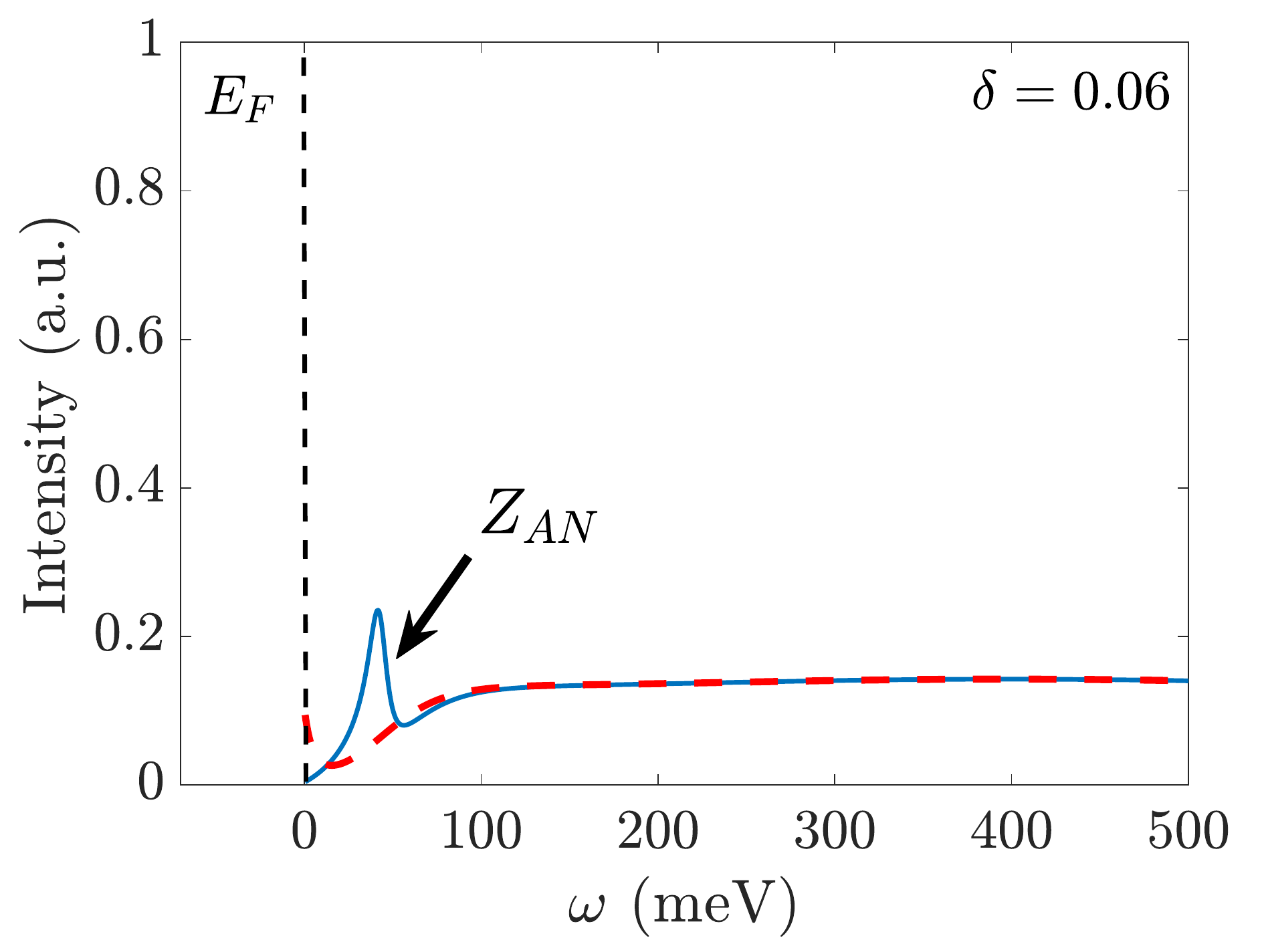}
\end{minipage}
\begin{minipage}[h]{0.33\textwidth}
\includegraphics[width=\textwidth]{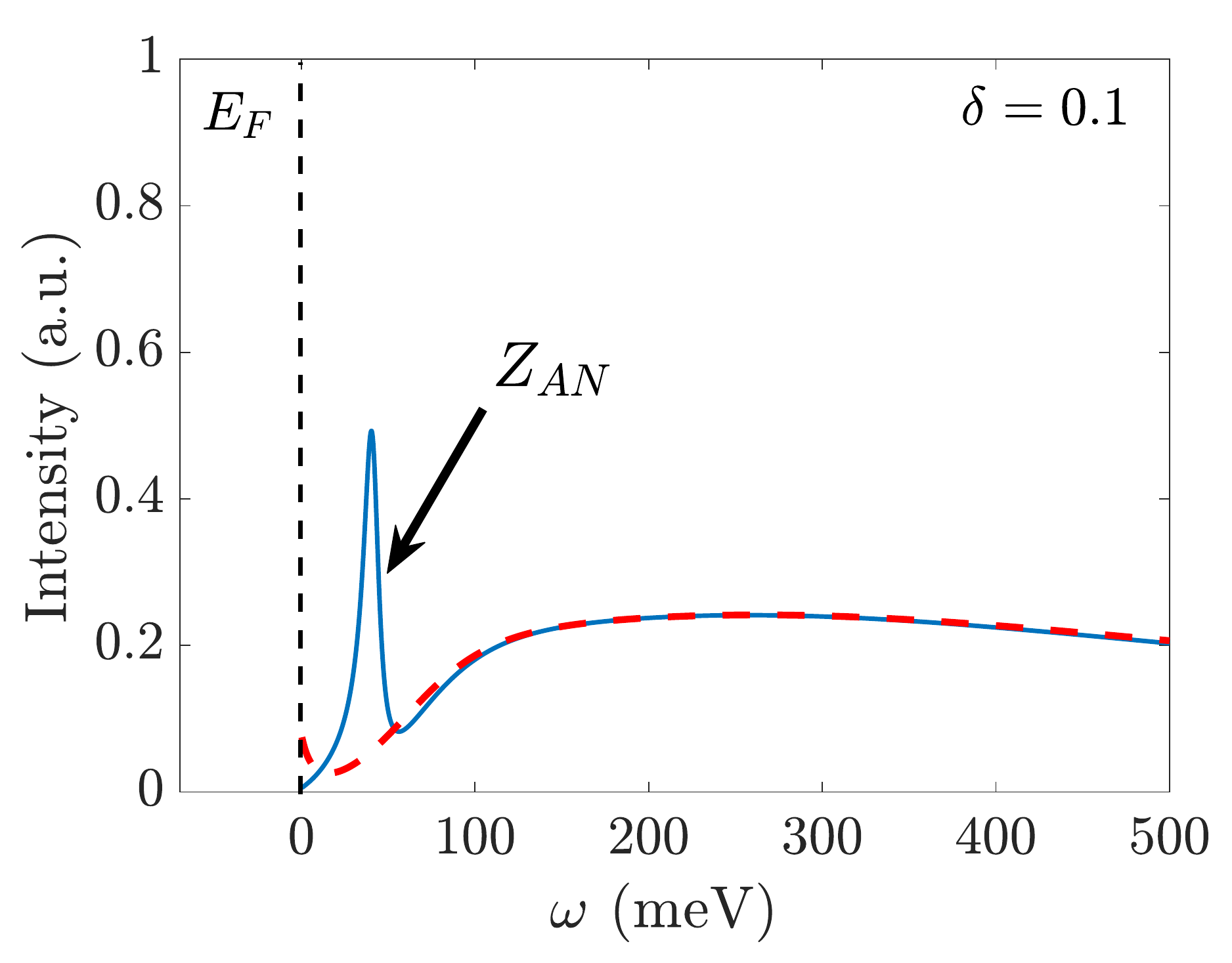}
\end{minipage}
\begin{minipage}[h]{0.316\textwidth}
\includegraphics[width=\textwidth]{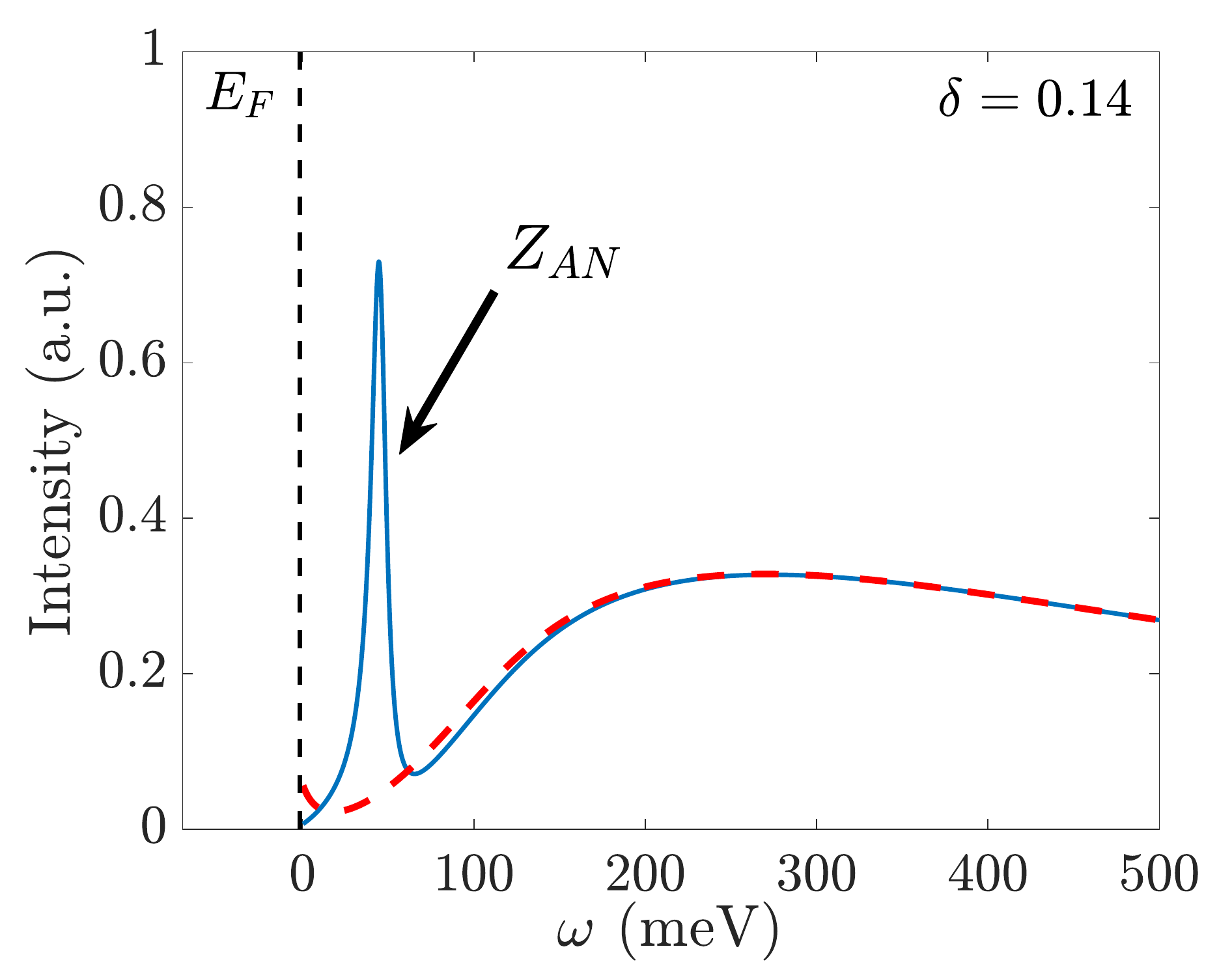}
\end{minipage}
\caption{The energy distribution curves (EDCs) near the antinodal regime of $(\pi,0)$ [cf. Fig. \ref{position}(a)] in the SC phase (blue) and LPP (red dashed) at $\delta= 0.06$, 0.1, and 0.14. The quasiparticle weight $Z_{\text{AN}}$ is only present in the SC phase, giving rise to a ``peak-dip-hump'' feature, whereas it diminishes in the LPP in contrast to a finite $Z_{\text{N}}$ coinciding with the Fermi arc in the nodal region.  Note that the ``hump'' feature is present in both the LPP and SC phase due to the fractionalized mode of the $a$-spinon excitation.}
\label{EDC}
\end{figure*}

Finally, corresponding to the spectral weight $Z_{\text{AN}}$ at a momentum near the neighborhood of the momentum $(\pi,0)$ in the antinodal regime [Fig. \ref{position}(a)], the energy distribution curve (EDC) is shown in Fig. \ref{EDC} at various dopings. A ``peak-dip-hump'' structure is generally present in Fig. \ref{EDC}. Here the ``peak'' is attributed to the Bogoliubov quasiparticle in the anti-nodal regime, with its peak intensity $Z_{\text{AN}}$ monotonically dependent on the holon density or superfluid density as indicated in Fig. \ref{position}(c). On the other hand, ``hump'' feature should be solely attributed to the high-energy fractionalized mode related to the gapped $a$-spinon excitation in $\hat{D}_0$, similar to that in the LPP as indicated in Figs. \ref{LPP scanning} and \ref{position}(a), where $Z_{\text{AN}}$ is absent, while the Fermi arc is characterized by the spectral weight $Z_{\text{N}}$ which remains finite. The ``hump'' structure is  explicitly shown in the high-energy regimes of Fig. \ref{EDC} , which is further broadened via $f_0$ in $\hat{\cal D}_0$ phenomenologically (cf. Appendix \ref{App.B}).

\section{Summary\label{section5}}
In this work, we have studied the spectral function based on a single-particle Green's function constructed in the phase-string formulation of the $t$-$J$ model. We have shown that the basic characteristics of the phase-string-induced fractionalization can be directly exhibited in the spectral function and probed by the ARPES experiment. For example, the Fermi arc in the lower pseudogap phase may be understood as the fractionalization of the quasiparticle excitation \textit{outside} the arc, while it remains coherent \textit{inside} the arc. In the superconducting phase, a superconducting $d$-wave gap is opened up along the Fermi arc. Such a conventional Bogoliubov quasiparticle can further connect to an emergent sharp mode outside the Fermi arc with a distinct (larger) gap extrapolated to the antinodal region, lead- ing to a two-gap structure. Similarly, a ``kink'' in the quasiparticle spectrum along the nodal direction has been also consistently found. In particular, a systematic doping dependence of these novel features has been deter- mined. The overall agreement between the theory and the ARPES experiments are remarkable.

Such a phenomenological Green's function is composed of two components, $\hat{G}_0$ and $\hat{\cal D}_0$ in Eq. (\ref{G3}), to characterize the dichotomy between a coherent quasiparticle and the fractionalization. Here $\hat{\cal D}_0$ describes the propagation of the fractionalized particles in the lower pseudogap and superconducting phases, while $\hat{G}_0$ depicts the propagation of an injected hole as a coherent quasiparticle in a renormalized mean-field description \cite{ZhangFC1988} before decaying into a fractionalized state or after a recombination from the fractionalization. In contrast to a large Fermi surface in $\hat{G}_0$ satisfying the Luttinger volume of the total electrons, $\hat{\cal D}_0\simeq \hat{D}_0$ is essentially characterized by four small Fermi pockets 
of an area proportional to the doping concentration $\delta$ as illustrated in Fig. \ref{Fermi pocket}.

Even though the electrons are fully fractionalized in the ground state [Eq. (\ref{ground state wavefunction})] of the lower pseudogap and superconducting phases \cite{Weng2011,MaYao2014}, a quasiparticle as an excitation can still emerge as a recombination of the fractionalized particles within the gap of the fractionalized $a$-spinon (cf. Fig. \ref{LPP scanning}). Besides the $s$-wave-like gap of the $a$-spinon, the Fermi pocket of the $a$-spinon further provides a peculiar protection of the stability and coherence of the quasiparticle excitation. This is because any decay of a quasiparticle into an $a$-spinon via Eq. (\ref{fractionalization}) within the Fermi pocket area would violate the Pauli exclusion principle. It thus leads to the Fermi arcs in the lower pseudogap phase at low doping. The overall excellent agreement with the experiment as a function of doping is simply due to the fact that the size of the $a$-pocket is solely determined by $\delta$, which is independent of the mean-field approximation \cite{MaYao2014} in treating $\hat{D}_0$ in Eq. (\ref{fractionalized}).

Therefore, we have shown that the unique fractionalization in the $t$-$J$ model may provide a unified and consistent description of the ARPES experimental results in the cuprate. Or, in other words, an ARPES measurement may effectively reveal the electron fractionalization in a doped Mott insulator. A doped hole is fractionalized into a composite structure with emergent internal degrees of freedom in Eq. (\ref{fractionalization}). In contrast to the full spin-charge separation in the one-dimensional case, here the holon in 2D is generally accompanied by a spin (specified by the $a$-spinon) to form a composite with internal degrees of freedom and a spatial size comparable to that of an RVB pair in the background \cite{Weng2011}, which is always finite in the pseudogap/superconducting phase \cite{Weng2011,MaYao2014} such that the recombination into a quasiparticle occurs by a finite probability $\propto |\lambda|^2$. By contrast, $\lambda\rightarrow0$ for a full separation of spin and charge in the long-range antiferromagnetic order regime, where the Fermi pockets collapse into four Fermi points at $\bs{k}_0=(\pm\pi/2,\pm\pi/2)$ with a vanishing quasiparticle spectral weight in the thermodynamic limit, which has been recently shown \cite{SingleHole} as the precursor of the Fermi arc phenomenon in the one-hole-doped limit of the ground state (\ref{ground state wavefunction}).

\begin{acknowledgments}
Stimulating discussions with Shuai Chen, Dong-Lai Feng, Tai-Kai Ng, Yi-Zhuang You, Shuo Yang, Yang Qi, Zheng-Cheng Gu, Yayu Wang, Ziqiang Wang, and Xing-Jiang Zhou are acknowledged. This work is partially supported by Natural Science Foundation of China (Grant No. 11534007), MOST of China (Grant Nos. 2015CB921000 and 2017YFA0302902). YM acknowledges the support from the NSFC via projects 11704308. \\
\end{acknowledgments}

\newpage

\appendix

\section{Effective Hamiltonian of the microscopic theory\label{App.A}}

The ground state in Eq. (\ref{ground state wavefunction}) is the direct product of $|\Phi_h\rangle$, $|\Phi_a\rangle$, and $|\mathrm{RVB}\rangle$ as the mean-field solutions of $H_h$, $H_a$ and $H_s$, respectively, in the effective Hamiltonian
\begin{align}
H_{\text{eff}}=H_h+H_a+H_s,
\end{align}
\begin{widetext}
\begin{align}
\begin{aligned}
&H_h=-t_h\sum\limits_{\langle i,j\rangle}h_i^\dag h_je^{i\left(A_{ij}^s+eA_{ij}^e\right)}+h.c.+\lambda_h\left(\sum\limits_{i}h_i^\dag h_j-\delta N\right)\\
&H_a=-t_a\sum\limits_{\langle i,j\rangle,\sigma}\sigma a_{i\sigma}^\dag a_{j\sigma}e^{-i\phi_{ij}^0}+h.c.-\gamma\sum\limits_{\langle i,j\rangle}\left(\hat{\Delta}_{ij}^a\right)^\dag\hat{\Delta}_{ij}^a+\lambda_a\left(\sum\limits_{i,\sigma}a_{i\sigma}^\dag a_{i\sigma}-\delta N\right)\\
&H_s=-J_s\sum\limits_{\langle i,j\rangle}\hat{\Delta}_{ij}^s+h.c.+\lambda_b\left(\sum\limits_{i,\sigma}b_{i\sigma}^\dag b_{i\sigma}-N\right)
\end{aligned}
\label{mean-field}
\end{align}
\end{widetext}
which are deduced from the fractionalized representation \cite{Weng2011,MaYao2014} of the $t$-$J$ Hamiltonian.

In $H_h$, the holon carries the full electric charge $+e$ coupling to the external electromagnetic field $A_{ij}^e$ as well as the internal link variable
\begin{align}
A_{ij}^s=\frac{1}{2}\sum\limits_{l\ne i,j}[\theta_i(l)-\theta_j(l)](n_{l\uparrow}^b-n_{l\downarrow}^b)
\label{As}
\end{align}
originated from the phase string effect. Since $A_{ij}^s\simeq0$ as the background $b$-spinons are short-range RVB-paired, generally one expects the \emph{holon condensation} at low temperatures, which defines the LPP and SC phase. The LPP corresponds to the appearence of \textit{free vortices} in $A_{ij}^s$ without destroying the Bose condensation. The $a$-spinon in $H_a$ is gauge neutral as ``protected'' by the ODLRO: ${\Delta}^a= \langle \hat{\Delta}_{ij}^a\rangle \neq 0$, which has been found to be $s$-wave \cite{MaYao2014} to determine a BCS-like $|\Phi_a\rangle$, where
\begin{align}
\hat{\Delta}_{ij}^a=\sum\limits_{\sigma}\sigma a_{i\sigma}^\dag a_{j\bar\sigma}^\dag e^{-i\phi_{ij}^0}
\label{DeltaA}
\end{align}
which is independent of the gauge choice of $\phi_{ij}^0$. Here $\phi_{ij}^0$ describes the background $\pi$-flux per plaquette:
\begin{align}
\phi_{ij}^0=\frac{1}{2}\sum\limits_{l\ne i,j}[\theta_{i}(l)-\theta_j(l)]
\label{pi-flux}
\end{align}
The $b$-spinon state  $|\mathrm{RVB}\rangle=\mathcal{P}|\Phi_b\rangle$, with $|\Phi_b\rangle$ determined by $H_s$, which is underpinned by the order parameter $\Delta^s=\langle\hat{\Delta}_{ij}^s\rangle$ where
\begin{align}
\hat\Delta_{ij}^s=\sum\limits_\sigma e^{-i\sigma A_{ij}^h}b_{i\sigma}b_{j\bar\sigma}
\label{DeltaS}
\end{align}
with the link variable
\begin{align}
A_{ij}^h=\frac{1}{2}\sum\limits_{l\ne i,j}[\theta_i(l)-\theta_j(l)]n_{l}^h
\label{Ah}
\end{align}
Note that $A_{ij}^h$ can be treated as a uniform flux in the LPP and SC phase as the holons are condensed. But $\Delta^s\ne0$ will further extend to characterize a short-range RVB state known as the upper pseudogap phase (UPP) up to $T_0$ as shown in Fig. \ref{phase diagram}.

\begin{figure}
\centering
\includegraphics[width=0.48\textwidth]{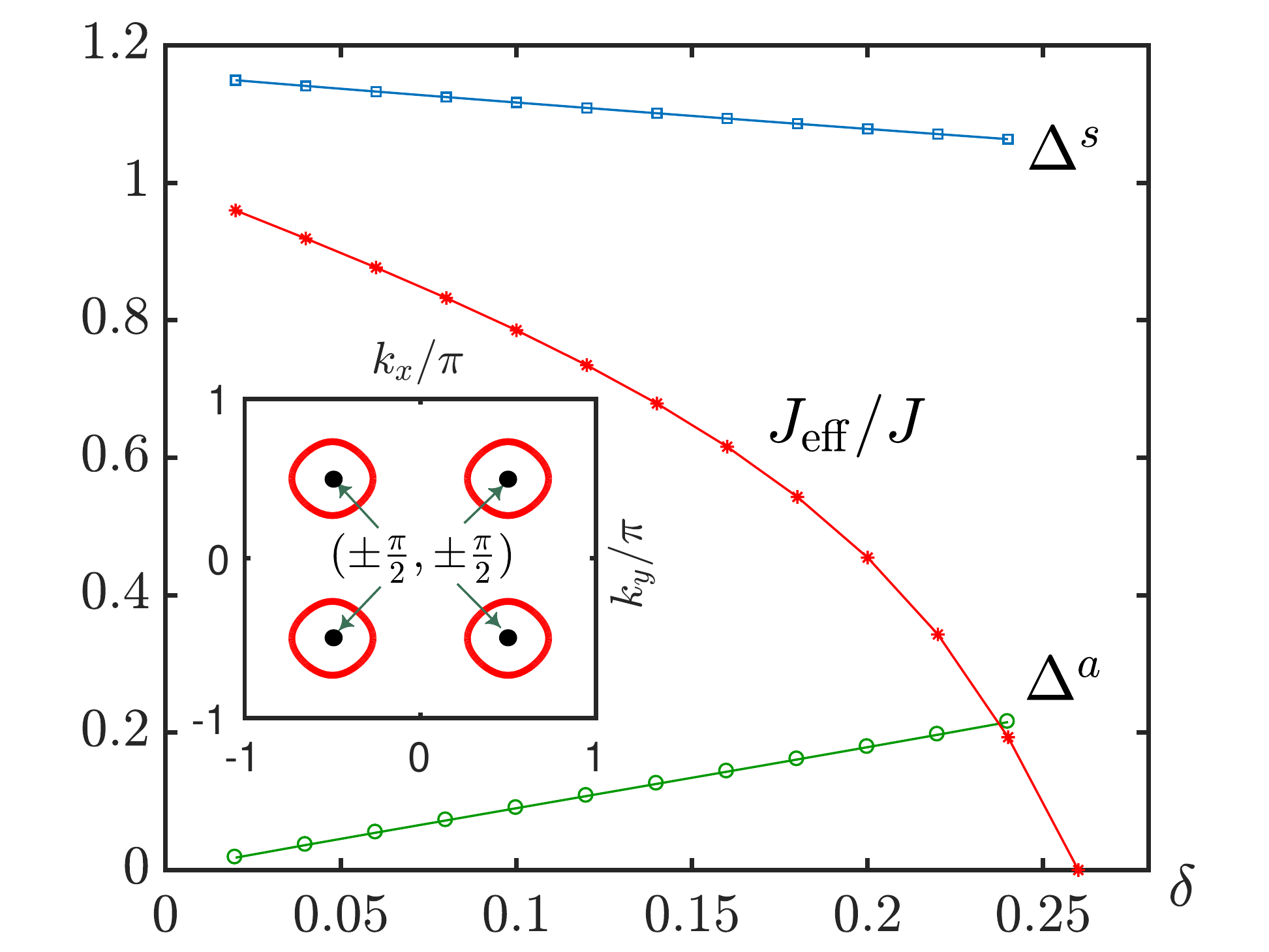}
\caption{Two-component RVB order parameters, $\Delta^s$ for $b$-spinons and $\Delta^a$ for $a$-spinons, as well as the effective superexchange coupling $J_{\text{eff}}/J$ based on the mean-field theory \cite{MaYao2014}. Insert: the lowest energy contours of the $a$-spinons with $s$-wave pairing $\Delta^a$, which are characterized by four Fermi pockets located at $\bs{k}_0=(\pm\pi/2,\pm\pi/2)$ in the fractionalized propagator [Eq. (\ref{G11})].}
\label{order parameter}
\end{figure}

The Lagrangian multipliers in Eq. (\ref{mean-field}) are introduced to enforce the constraints under the projection $\mathcal{P}$ \cite{MaYao2014}. Based on the self-consistent calculations, the following parameters as a function of doping concentrations in the ground state can be numerically determined \cite{MaYao2014}:
\[\Delta^s,~\Delta^a,~\chi^a,~\lambda_a,~\lambda_b,~\gamma\]
together with:
\begin{align}
\left\{
\begin{aligned}
&J_{\text{eff}}=J(1-\delta)^2-2\gamma\delta^2\\
&J_s=J_{\text{eff}}\Delta^s/2
\end{aligned}
\right.
\label{Jeff}
\end{align}
The doping dependent $b$-RVB order parameter $\Delta^s$ and $s$-wave pairing order parameter $\Delta^a$ for $a$-spinons, and the effective coupling $J_{\text{eff}}/J$ are shown in the main panel of Fig. \ref{order parameter} under the choice of $t_a=2J$ \cite{MaYao2014}, which constitutes the basic parameters in the study of the single-particle spectral function.

\section{Phase fluctuation in $\hat{\cal D}_0$: A phenomenological description of $f_0$\label{App.B}}

Note that in general, the single-particle propagator in Eq. (\ref{G2}) involves a phase correlation function $f_0$ defined in Eq. (\ref{f0}). Here $f_0$ describes the vortices (antivortices) attached to the background $b$-spinons whose fluctuations will give rise to the broadenings in both momentum and frequency of the Green's function according to its definition in Eq. (\ref{f0}). But we expect that the essential low-energy features of Eq. (\ref{G3}), i.e., the Fermi arc in the LPP and two-gap structure in the SC phase, should not be changed by the additional broadening effect introduced by $f_0$ because of the gapped nature of the $a$-spinons. In the following, we shall make some simple assumptions on the form of $f_0$. 

Due to the short-range RVB pairing of the $b$-spinons, the majority of vortices and antivortices are tightly paired to make the phase coherence in the SC phase, while a finite density of free vortices appear in the LPP to render the SC order parameter $\Delta^{\text{SC}}=0$, which is also expected to lead to a spatial decay of the phase correlation function $f_0$. One may phenomenologically introduce an approximate form to characterize the phase fluctuation by
\begin{align}
f_0(i,j;\tau)\Rightarrow e^{-|\bs{r}_i-\bs{r}_j|/b}e^{\tau/a}
\label{broaden1}
\end{align}
where $\tau$ is the imaginary time, $a$ and $b$ are spatial and temporal correlation length, respectively. In the following calculation, we shall choose $a=1.2\mathrm{(meV)}^{-1}$ and $b=4a_0$, but the main ARPES features will not be sensitive to the choice of the parameters made here, which are introduced mainly to smooth the spectral function in Sec. \ref{section4}. In particular, the phase factor $f_0$ will also generally contribution to the decay with the real time (i.e., $a$ should be complex) at higher energies. For the convenience of analysis, one may incorporate such an effect in calculating the convolution of $\hat{\cal D}_0$ in Eq. (\ref{G4}) in the frequency space by introducing a Lorentzian broadening:
\begin{align}
\frac{\eta}{\omega^2+\eta^2}
\label{Lorentzian}
\end{align}
where the width of the Lorentzian broadening is taken as $\eta=30$ meV throughout for simplicity.
Such a broadening originally comes from the convolution with the phase fluctuations in Eq. (\ref{broaden1}) after an analytic continuation to the real time/real frequency axis, where the vortices (antivortices) attached to the $b$-spinons will get significantly excited with an energy higher than the gap of $a$-spinons (Ref. \onlinecite{Weng2011,MaYao2014}). We emphasize that such a phenomenological assumption does not affect the main features near the Fermi energy, but mainly on the broadness of the high energy part of the EDC (cf. Sec. \ref{4C}). 

\bibliography{citation}

\end{document}